\documentclass[aip,jcp,superscriptaddress,showpacs,showkeys,reprint]{revtex4-1}

\pdfoutput=1 %arxiv

\usepackage{amssymb}
\usepackage{amsmath}
\usepackage{graphicx}
\usepackage{hyperref}   % use for hypertext links, including those to external documents and URLs
\usepackage{color}      % use if color is used in text

\newcommand{\vect}[1]{\boldsymbol{\mathbf{#1}}}

\newcommand{\homega}{\hat{\omega}}

\newcommand*{\citen}[1]{%
  \begingroup
    \romannumeral-`\x % remove space at the beginning of \setcitestyle
    \setcitestyle{numbers}%
    \cite{#1}%
  \endgroup   
}

\begin{document}

\title{Phase diagrams of charged colloidal rods: can a uniaxial charge distribution break chiral symmetry?}

\author{Tara Drwenski}
\email{t.m.drwenski@uu.nl}
\affiliation{Institute for Theoretical Physics, Center for Extreme Matter and Emergent Phenomena, Utrecht University, Leuvenlaan 4, 3584 CE Utrecht, The Netherlands}

\author{Simone Dussi} 
\affiliation{Soft Condensed Matter, Debye Institute for Nanomaterials Science, Utrecht University, Princetonplein 5,
3584 CC Utrecht, The Netherlands}

\author{Michiel Hermes}
\affiliation{School of Physics and Astronomy, The University of Edinburgh, King's Buildings, Peter Guthrie Tait Road, Edinburgh, EH9 3FD, United Kingdom}

\author{Marjolein Dijkstra} 
\affiliation{Soft Condensed Matter, Debye Institute for Nanomaterials Science, Utrecht University, Princetonplein 5,
3584 CC Utrecht, The Netherlands}

\author{Ren\'{e} van Roij} 
\email{r.vanroij@uu.nl}
\affiliation{Institute for Theoretical Physics, Utrecht University, Leuvenlaan 4, 3584 CE Utrecht, The Netherlands}

\date{\today}

\begin{abstract}
      We construct phase diagrams for charged rodlike colloids within the second-virial approximation as a function of rod concentration, salt concentration, and colloidal charge. Besides the expected isotropic-nematic transition, we also find parameter regimes with a coexistence between a nematic and a second, more highly aligned nematic phase including an isotropic-nematic-nematic triple point and a nematic-nematic critical point, which can all be explained in terms of the twisting effect. We compute the Frank elastic constants to see if the twist elastic constant can become negative, which would indicate the possibility of a cholesteric phase spontaneously forming. Although the twisting effect reduces the twist elastic constant, we find that it always remains positive. In addition, we find that for finite aspect-ratio rods the twist elastic constant is also always positive, such that there is no evidence of chiral symmetry breaking due to a uniaxial charge distribution.
\end{abstract}


\maketitle

%----------------------------------------------------------------------------------------
%	Section 1
%----------------------------------------------------------------------------------------
\section{Introduction}

The isotropic-nematic phase transition in dispersions of rigid, rodlike colloids occurs at sufficiently high concentration of rods. For uncharged rods, this phase transition is purely the result of a competition between orientational entropy, which is maximized in the isotropic phase, and the translational entropy, which favors the nematic phase, where rods tend to align along a nematic director $\hat{n}$. For long, rigid, needle-like rods, this phase transition is accurately described by Onsager's second-virial theory.~\cite{Onsager}

Many experimental systems do not form ordinary nematic phases, but instead form a cholesteric (chiral nematic) phase, where the nematic director field has a helical arrangement with a pitch much larger than the colloidal dimensions. Though the cholesteric phase is ubiquitous in experimental systems, the relationship between particle properties and macroscopic chirality remains unclear.~\cite{Harris1999,Grelet2003} An illustrative example of this involves suspensions of filamentous fd virus, which are semi-flexible charged need\-les with a chiral structure that form a cholesteric phase in a density regime that depends on the ionic strength. In Ref.~[\citen{Grelet2003}], however, fd-virus particles sterically stabilized by a coating with the neutral polymer polyethylene glycol (PEG) exhibited a phase diagram and a nematic order parameter independent of the ionic strength, but surprisingly, the fd-PEG continued to form a cholesteric phase with a pitch that did vary with the ionic strength. 

Furthermore, molecular chirality does not guarantee macroscopic chirality. For example, the virus Pf1, with a chiral structure very similar to that of fd, does not form a cholesteric phase (or its pitch is too large to observe experimentally).~\cite{Fraden2000} Indeed, subtle alterations of the surface properties of fd that do not have a large effect on the phase diagram can have an appreciable effect on the cholesteric pitch.~\cite{Zhang2013} Reversing the surface charge of fd from negative to positive even prevented the observation of a cholesteric pitch, though the chemical modification of fd may have also introduced additional attractive forces.~\cite{Zhang2013} The fact that the cholesteric pitch is very sensitive to particle surface properties was also shown in a study of M13, which is a charged, large-aspect ratio bacteriophage with a right-handed structure that is shown to form a left-handed macroscopic phase.~\cite{Tombolato2006} Though steric effects are shown to favor a right-handed phase, charges added along grooves of the coarse-grained representation of M13 caused the calculated pitch to become left-handed.~\cite{Tombolato2006} The microscopic origin of chirality in colloidal suspensions remains a mystery despite many interesting recent works,\cite{Wensink2015,Ferrarini2015,Dussi} though charge seems to be one of the crucial ingredients.~\cite{Grelet2003,Zhang2013,Tombolato2006,Wensink2009,Wensink2011} 

A wide variety of experimental systems that display nematic or cholesteric phases involve electrostatic interactions,~\cite{Vroege1992} for example, filamentous viruses,~\cite{Lapointe1973, Fraden1995, Dogic2006} actin filaments,~\cite{Suzuki1991,Coppin1992,Furukawa1993} cellulose derivatives,~\cite{Werbowyj1976, Werbowyj1980} and single-walled carbon nanotubes in superacids.~\cite{Davis2004, Rai2006} For strong electrostatic interactions or short screening lengths, the isotropic-nematic phase transition is well understood. Onsager~\cite{Onsager} was the first to note that the soft repulsion can be treated by renormalizing the diameter of the rods. Stroobants et al.~\cite{Lekkerkerker1986} showed that there is a second effect for strong electrostatic interactions, namely a ``twisting" due to the angle dependence of the electrostatic potential, which makes the rods resist aligning. 

Weakly charged rods have also been studied extensively. In Refs.~[\citen{Semenov1988, Nyrkova1997}], a scaling theory was used to give qualitative predictions for charged rods. Interestingly, in a certain region of low charge density and moderate screening they predict that a competition between steric and electrostatic effects leads to a coexistence between a nematic and a highly oriented nematic phase. Most predictions of Refs.~[\citen{Semenov1988, Nyrkova1997}], including the existence of the nematic-nematic coexistence were confirmed in Ref.~[\citen{Chen1996}], using a Debye-H\"{u}ckel-like theory that includes some many-rod correlations. Another interesting result is that the correlation electrostatic energy due to charge fluctuations in a many-body system of charged rods and counterions makes orientational order more favorable, stabilizes a weakly ordered nematic at small rod concentrations, and leads to the possibility of two nematic phases.~\cite{Potemkin2002,Potemkin2005}

The goal of this paper is to quantitatively examine both weak and strong electrostatic interactions using second-virial theory and additionally to investigate how charge affects the stability of the nematic phase with respect to spontaneous twist deformations. In Sec.~\ref{sect:Onsager}, we review second-virial theory for charged rods and extend previous results to weakly charged rods. We also determine for which parameters the twisting effect becomes important for weakly charged rods. In Sec.~\ref{sect:PhaseDiagrams}, we construct the phase diagrams and identify the parameter regime where nematic-nematic coexistence can occur. We then briefly discuss the possibility of seeing the nematic-nematic coexistence experimentally in Sec.~\ref{sect:experiment}. In Sec.~\ref{sect:ElasticConsts}, we investigate if the twisting effect can stabilize a cholesteric phase. We do this by calculating the Frank elastic constants of the nematic phase and examining the relationship between the twisting effect and the twist elastic constant. We are especially interested whether the twist elastic constant (evaluated in the nematic state) can become negative, which would indicate the possibility of a cholesteric phase spontaneously forming. Finally, we examine the sign of the twist elastic constant for finite aspect-ratio rods in Sec.~\ref{sect:Simone}. We end with a summary and conclusions in Sec.~\ref{sect:conclusion}.

%----------------------------------------------------------------------------------------
%	Section 2
%----------------------------------------------------------------------------------------
\section{Onsager theory}\label{sect:Onsager}

\subsection{Second-virial term}

	% ---------------
	%  FIG SPHEROCYLINDERS
	% ---------------
	\begin{figure}[htb]
        \includegraphics[width=\linewidth, bb = 0 0 350 400]{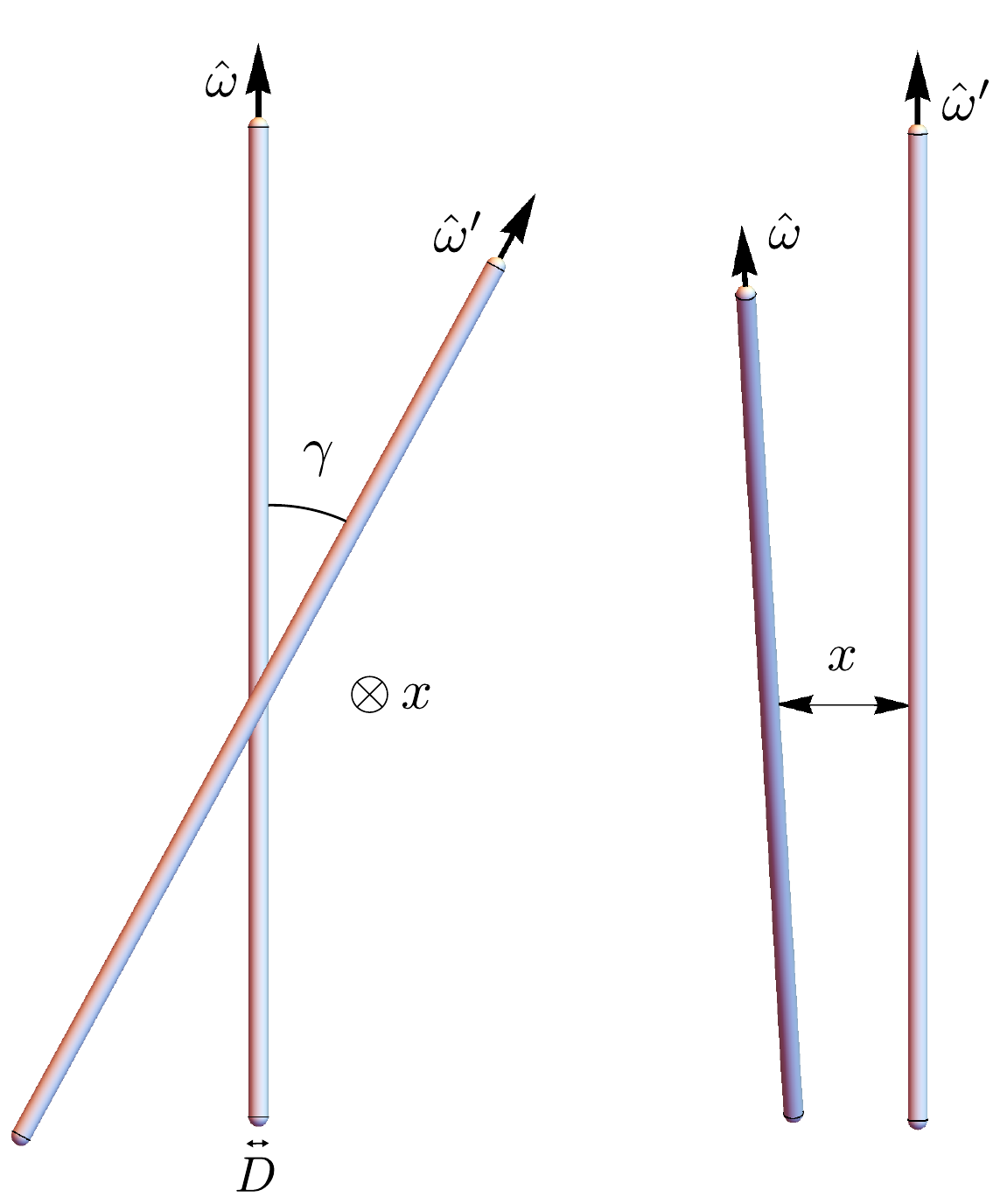}
        \caption{Illustration of two charged spherocylinders with diameters $D$ from two different viewpoints. The rods are oriented along unit vectors $\homega$ and $\homega'$, with $\gamma = \cos^{-1} \left( \homega \cdot \homega' \right)$  the angle between the rods and $x$ the shortest distance between them.}
        \label{fig:spherocylinders}
    \end{figure}

We consider $N$ charged colloidal rods of length $L$ and diameter $D$ suspended in an electrolytic solvent characterized by a salt concentration $\rho_s$ and a dielectric permittivity $\epsilon$. The system has a total volume $V$ and a temperature $T$. The Bjerrum length is given by $\lambda_B = e^2 \!/(4 \pi\, \epsilon_0 \, \epsilon \,k_\text{B} T)$, with $e$ the elementary charge, $k_\text{B}$ the Boltzmann constant, and $\epsilon_0$ the vacuum permittivity, and the Debye screening length is defined as $\kappa^{-1} = 1/\sqrt{8 \pi \lambda_B \rho_s}$.~\cite{mcquarrie} 

In addition to a hard-core repulsion between a pair of rods, there is also a screened electrostatic interaction, approximated by the interaction between two line charges with effective linear charge density $v_\text{eff} = Z/L$ with $Z$ the number of elementary charges on a rod. The form of this electrostatic interaction for infinitely long cylinders in the Debye-H\"{u}ckel approximation is well known.~\cite{Fixman, Brenner, Stigter} The pair potential $U(x, \gamma)$ is given by
	% ---------------
	\begin{equation}
	\label{eq:potential}
	 \beta  U(x, \gamma)= \left\{
	     \begin{array}{cl}
	       \infty, & x \leq D \\
	       \displaystyle\frac{\mathcal{A} \, e^{-\kappa x }}{\kappa D |\sin \gamma|}, & x>D ,
	     \end{array}
	   \right. 
	\end{equation}
	% ---------------
with $\beta =(k_\text{B} T)^{-1}$, $x$ the minimum separation between the two rods, $\gamma$ the angle between the rods with orientations $\homega$ and $\homega'$ defined by $\cos \gamma = \homega \cdot \homega'$ (see Fig.~\ref{fig:spherocylinders}), and where we have introduced the dimensionless coupling parameter
	% ---------------
	\begin{equation}
		\mathcal{A} = 2 \pi \, v_\text{eff}^2\, \lambda_B \, D.
		\label{eq:charge}
	\end{equation}
	% ---------------
 Eq.~(\ref{eq:potential}) is valid when $\kappa^{-1} \ll L$ and $x \ll L$. 

Following Onsager,~\cite{Onsager} we study the phase behavior of this suspension of charged needles in terms of the single-rod orientation distribution function $\psi(\hat{\omega})$, which suffices for translationally invariant phases. The distribution $\psi(\hat{\omega})$ is normalized as
	% ---------------
	\begin{equation}
	\label{eq:norm}
		\int \psi(\homega) \, d\homega = 1.
	\end{equation}
	% ---------------
Assuming $L \gg D$, we can write the Helmholtz free energy functional $F[\psi]$ of a suspension of rods in the second-virial approximation as
	% ---------------
	\begin{align}
	\label{eq:freeEnergy}
		\frac{\beta  F[\psi]}{ V} =&  \rho (\ln \mathcal{V}\rho -1)+ \rho \int \psi(\homega)\ln \psi(\homega) \, d\homega \nonumber\\
			&+\frac{1}{2} \rho^2 \iint  E(\homega,\homega') \psi(\homega) \psi(\homega') \, d\homega \,d\homega' \nonumber\\
			& + \mathcal{O}(\rho^3),
	\end{align}
	% ---------------
where $\rho = N/V$ is the number density and $\mathcal{V}$ is a thermal volume. In Eq.~(\ref{eq:freeEnergy}), the first term gives the translational entropy and the second gives the orientational entropy. The third term is the second-virial term, with the ``excluded volume" term $E (\homega, \homega') $ defined as
	% ---------------
	\begin{align}
	\label{eq:ExclVolMayer}
	E (\homega_1, \homega_2) &=	-\frac{1}{V} \iint \Phi(\mathbf{r}_1-\mathbf{r}_2;\homega_1,\homega_2) \, d\mathbf{r}_1 \, d\mathbf{r}_2 \nonumber \\
	&=-\int \left[ e^{-\beta U(\mathbf{r}_{12};\homega_1,\homega_2)}-1 \right]  \, d\mathbf{r}_{12} ,
	\end{align}
	% ---------------
where we have used translational invariance, defined $\mathbf{r}_{12} =\mathbf{r}_1-\mathbf{r}_2$, and introduced the Mayer function $\Phi = e^{-\beta U}-1$ which depends on $U(\mathbf{r}_{12};\homega_1,\homega_2)$, the pair potential between a rod with orientation $\homega_1$ and position $\vect{r}_1$ and a second rod with orientation $\homega_2$ and position $\vect{r}_2$. 

Now, performing the integration over the Mayer function in Eq.~(\ref{eq:ExclVolMayer}) with the potential given in Eq.~(\ref{eq:potential}) and using $d\vect{r}_{12} = L^2 |\sin \gamma| \, dx$, we can write $E(\gamma)= E(\homega, \homega')$ as~\cite{Fixman}
 	% ---------------
	\begin{align}
		\label{eq:ExclVol}
		E(\gamma) =& -2L^2 \, |\sin  \gamma| \int_0^\infty \Phi (x,\gamma)\, dx \nonumber \\
		=& 2L^2 D \, |\sin \gamma| \left\{ 1+\frac{1}{\kappa D}\left[  \text{ln}\left( \frac{A'}{|\sin \gamma|} \right) \right. \right. \nonumber\\
		 & \qquad \left. \left. + \gamma_E -\text{Ei}\left( -\frac{A'}{|\sin \gamma|}  \right)  \right]  \right\},
	\end{align}
 	% ---------------
where $\gamma_E \approx 0.5772$ is Euler's constant, the exponential integral Ei is defined as $\text{Ei}(y) = -\int_{-y}^\infty \exp(-t)/t \,dt$, and $A'=\mathcal{A} \, e^{-\kappa D}/(\kappa D).$ The function $\text{Ei}(-A')$ becomes negligible for $A' \gtrsim 2$, which is the approximation used in Ref.~[\citen{Lekkerkerker1986}]. In the present work, we also consider $A' \lesssim 2$, and hence we keep the $\text{Ei}$ term in Eq.~(\ref{eq:ExclVol}) throughout.

	% ---------------
	%  FIG
	% ---------------
	\begin{figure*}[ht!]
			\includegraphics[width=\linewidth, bb=0 0 482 170]{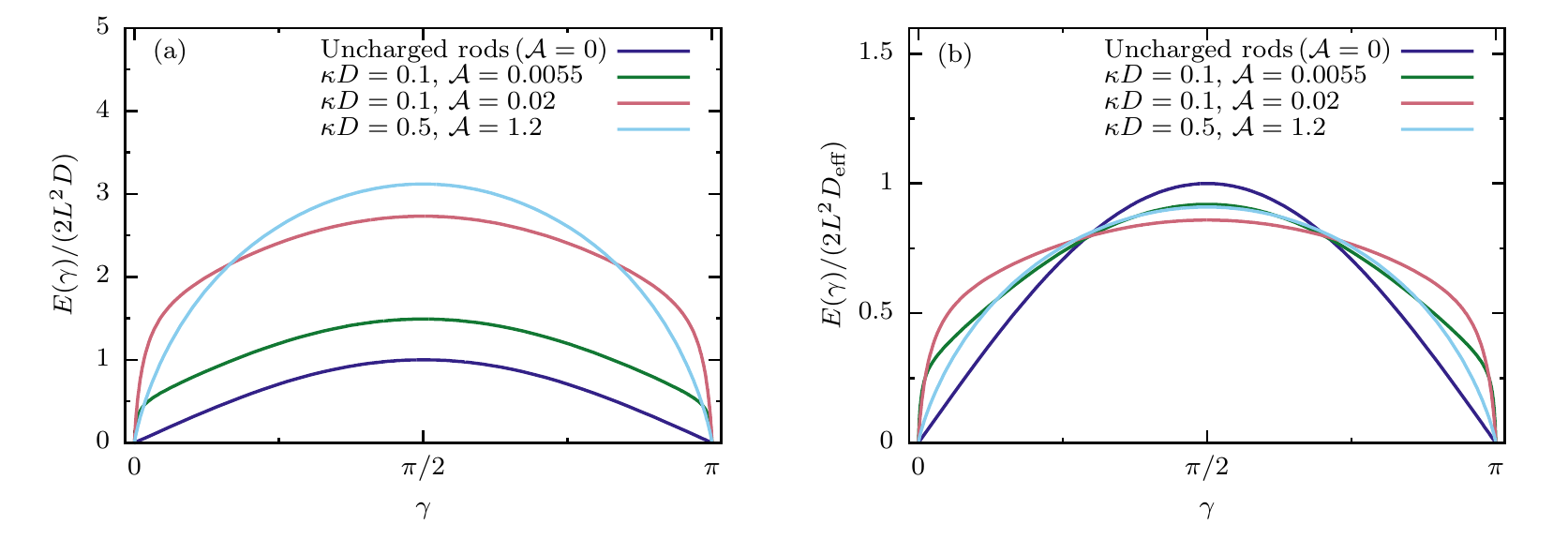}
                \caption{Dependence of ``excluded volume" $E(\gamma)$ (Eq.~\eqref{eq:ExclVol}) on the angle $\gamma$ between two rods (see Fig.~\ref{fig:spherocylinders}) for uncharged rods ($\mathcal{A}=0$) and charged rods for different values of screening parameter $\kappa D$ and Coulomb coupling $\mathcal{A}$. In (a), $E(\gamma)$ is scaled by the volume factor $2L^2D$. In (b), $E(\gamma)$ is scaled by effective volume factor $2L^2D_\text{eff}$ (see Eq.~\eqref{eq:Deff}). The value of twisting parameter $H$ (Eq.~\eqref{eq:newTwist}) is $0$, $0.71$, $1.1$, and $0.54$ for purple, green, pink, and light blue curves, respectively.}\label{fig:exclVol}
    \end{figure*}

The function $E(\gamma)$ of Eq.~(\ref{eq:ExclVol}) depends on the intrinsic excluded volume $L^2 D$ of the rods, the screening parameter $\kappa D$, and the parameter $A'$. However, in order to be able to vary the charge density of the needles and the salt concentration independently, we prefer to use $\mathcal{A}$ rather than $A'$ as an independent parameter, since $\mathcal{A}$ only depends on the charge of the rods (and the Bjerrum length) and not on $\kappa D$. In Fig.~\ref{fig:exclVol}(a), we plot $E(\gamma)$ as a function of the angle $\gamma$ between the rods for a few values of $\kappa D$ and $\mathcal{A}$, along with the hard-rod excluded volume for comparison. We observe essentially two effects compared to the hard-rod excluded volume, for which $E(\gamma)/(2 L^2 D)$ reduces to $|\sin\gamma| $. First, due to the charge there is an overall increase in the ``excluded volume" for all parameters, and second, there is a change in the shape of $E(\gamma)$ from that of hard rods, in particular the $\gamma$-dependence is much stronger at small $\gamma$'s, i.e. charged rods disfavor small angles much more than hard rods do. The former we describe by an increasing effective diameter $D_\text{eff}$ and the latter we describe by a ``twisting" parameter. These two effects will be discussed in Sec.~\ref{sect:Deff} and Sec.~\ref{sect:twist}, respectively.

The equilibrium orientation distribution function is obtained by minimizing $F[\psi]/V$ with respect to $\psi (\homega)$, at fixed $\rho$, $T$, $\kappa^{-1}$, and $\mathcal{A}$, which gives the integral equation~\cite{Vroege1992}
	% ---------------
	\begin{equation}
		\label{eq:EL}
		\ln \psi(\homega)+ \rho \int E(\gamma(\homega,\homega')) \psi(\homega') \, d\homega' = C,
	\end{equation}
	% ---------------
 with $C$ a constant that ensures that the constraint of Eq.~(\ref{eq:norm}) is satisfied. There is an analytic solution to Eq.~(\ref{eq:EL}), namely $\psi_\text{i}(\homega)=1/(4\pi)$, describing the isotropic phase which is the only (stable) one at sufficiently low $\rho$.~\cite{Onsager, Kayser1978, Mulder1989}

 At higher densities, $E(\gamma)$ becomes more important in Eq.~(\ref{eq:EL}) and the rods favor the nematic phase, where the orientation distribution function becomes peaked around a nematic director $\hat{n}$. We choose a coordinate system with the $z$-axis parallel to $\hat{n}$. The unit vector $\homega$ can be written as $\homega=(\sin \theta \, \cos \varphi, \sin \theta \, \sin \varphi, \cos\theta)$, where $\varphi$ is the azimuthal angle and $\theta$ is the polar angle with respect to $\hat{z}$. The orientation distribution function is independent of the azimuthal angle $\varphi$, has up-down symmetry, and hence we can write $\psi(\homega)=\psi(\homega \cdot \hat{n}) = \psi(\homega \cdot -\hat{n})$. To determine the orientation distribution function $\psi(\homega)$ for the nematic phase, we solve Eq.~(\ref{eq:EL}) using an iterative scheme on a discrete grid of polar angles $\theta \in [0,\pi/2)$.~\cite{Herzfeld1984, vanRoij2005}

% -------------------------
% 	Subsection 1
% -------------------------
\subsection{Effective diameter}\label{sect:Deff}

 We introduce the double orientational average in the isotropic phase $\langle \langle \cdot \rangle \rangle_\text{i}$, as 
	% ---------------
	\begin{equation}
		 \langle \langle  f(\homega, \homega') \rangle \rangle_\text{i} = \frac{1}{16\pi^2}\iint f(\homega, \homega')  \, d\homega \, d\homega',
	\end{equation}
	% ---------------
for an arbitrary function $f(\homega, \homega')$. We now follow Ref.~[\citen{Lekkerkerker1986}] and define
 	% ---------------
 	\begin{equation}
 	\label{eq:Deff}
 		D_\text{eff} = D+ \alpha \kappa^{-1} ,
 	\end{equation}
 	% ---------------
 with the effective double-layer thickness parameter
 	% ---------------
 	\begin{equation}
 		\label{eq:alpha}
 		\alpha =   \ln A' +\gamma_E + \ln 2 - \frac{1}{2}  - \frac{4}{\pi} \langle \langle |\sin \gamma| \, \text{Ei}\left( -\frac{A'}{|\sin \gamma|}  \right) \rangle \rangle_\text{i}  .
 	\end{equation}
 	% ---------------
 	One checks from Eq.~(\ref{eq:ExclVol}) that the second-virial coefficient in the isotropic phase can be written as
 	% ---------------
 	\begin{equation}
 	\label{eq:EIso}
 		\frac{1}{2}\langle \langle E(\homega, \homega') \rangle \rangle_\text{i} =\frac{\pi}{4}L^2 D_\text{eff},
 	\end{equation}
 	% ---------------
 where we have used
	% ---------------
	\begin{align}
		&\langle \langle| \sin  \gamma| \rangle \rangle_\text{i} = \frac{\pi}{4},\nonumber \\
		&\langle \langle -|\sin \gamma| \, \ln |\sin \gamma| \rangle \rangle_\text{i} = \frac{\pi}{4}\left(\ln 2- \frac{1}{2}\right).
	\end{align}
	% ---------------

Eq.~(\ref{eq:EIso}) is precisely the second-virial coefficient of uncharged rods with a diameter $D_\text{eff}$ (in the isotropic phase). This justifies the interpretation of $D_\text{eff}$ as the effective diameter of the charged needles. The parameter $\alpha$ (Eq.~(\ref{eq:alpha})), which vanishes for $\mathcal{A}=0$, is a result of the electrostatic repulsions, which effectively increase the diameter of the rods.~\cite{Onsager, Lekkerkerker1986} The term in Eq.~(\ref{eq:alpha}) involving the exponential integral has to be integrated numerically. In Fig.~\ref{fig:exclVol}(b), we plot $E(\gamma)$ scaled by $2L^2D_\text{eff}$ in order to emphasize the twisting effect, which is discussed in detail in the following section.

 	% ---------------
	%  FIG
	% ---------------
	 	\begin{figure*}[!htb] 
			\includegraphics[width=\linewidth, bb= 0 0 482 170]{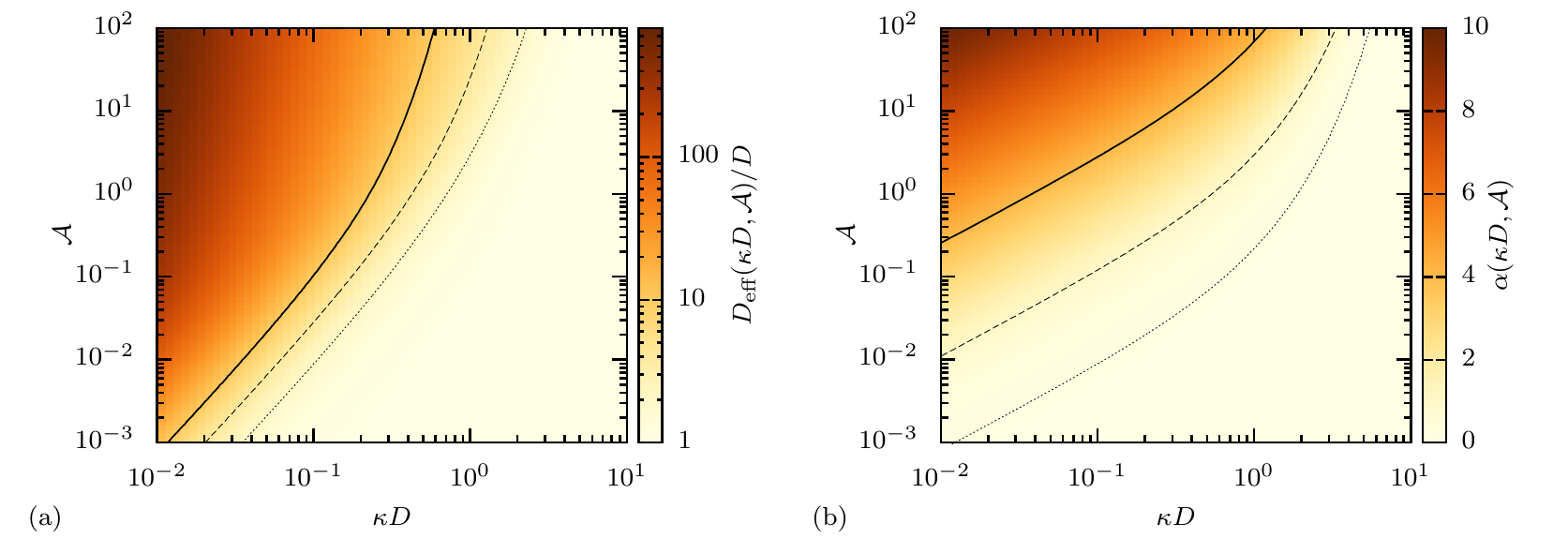}
        	\caption{Color plots indicating the value of (a) scaled effective diameter $D_\text{eff}/D$ (see Eq.~\eqref{eq:Deff}) with contours showing $D_\text{eff}/D = 2,4,10$ (dotted, dashed, solid respectively) and (b) effective double-layer thickness parameter $\alpha$ (see Eq.~\eqref{eq:alpha}) with contours showing $\alpha = 0.1,1,4$ (dotted, dashed, solid respectively) as a function of salt concentration $\kappa D$ and charge $\mathcal{A}$, on a log-log scale. Note that the color bar of (a) is in a log scale.}\label{fig:contourDeff}
     \end{figure*}

In Fig.~\ref{fig:contourDeff}, we present color plots as a function of $\kappa D$ and $\mathcal{A}$ indicating the value of (a) $D_\text{eff}$ (defined in Eq.~\eqref{eq:Deff}) and (b) of effective double-layer thickness parameter $\alpha$ (defined in Eq.~\eqref{eq:alpha}). We find that $D_\text{eff} \gg D$ (and so $\alpha > 0$) in a well-defined regime of sufficiently high $\mathcal{A}$ and low $\kappa D$, whereas $D_\text{eff} \approx D$ (thus $\alpha \approx 0$) in the complementary region. For example, fd virus is strongly charged ($v_\text{eff} \geq 4$ $e^-/$nm i.e. $\mathcal{A} \geq 500$) with a diameter of $D=6.6$ nm.~\cite{Purdy2004} The effective diameter of fd virus varies from $D_\text{eff}/D \approx 1.0$ at high ionic strength $\kappa D = 10$ (and so $\alpha \approx 0$) to $D_\text{eff}/D \approx 15$ at low ionic strength $\kappa D = 0.1$ (and so $\alpha \approx 1.4$).

% -------------------------
% 	Subsection 2
% -------------------------
\subsection{Twisting effect}\label{sect:twist}

In addition to an increase in the effective diameter, there is a second effect due to electrostatic interactions. This is a ``twisting" effect that is a result of the $|\sin \gamma|^{-1}$ term in the electrostatic potential (Eq.~\eqref{eq:potential}), first noted in Ref.~[\citen{Lekkerkerker1986}]. While the increase in the effective diameter of the rods tends to stabilize the nematic phase, this twisting effect tends to destabilize the nematic phase, pushing the isotropic-nematic phase transition to higher concentrations. This can be qualitatively understood if we consider $E(\gamma)$ in units of $2L^2D_\text{eff}$ as plotted in Fig.~\ref{fig:exclVol}(b), which reveals a strong $\gamma$ dependence for small angles $\gamma$.

In order to describe the twisting effect quantitatively, we follow Ref.~[\citen{Lekkerkerker1986}] and define the parameter
	% ---------------
	\begin{equation}\label{eq:oldTwist}
		h= \frac{1}{\kappa D_\text{eff}},
	\end{equation}
	% ---------------
such that Eq.~(\ref{eq:ExclVol}) can be rewritten as
	% ---------------
	\begin{align}
		\label{eq:exclVolNem}
		E(\gamma) &= 2L^2 D_\text{eff} \, |\sin \gamma| \nonumber \\
		&\times \left\{ \vphantom{\frac{A'}{|\sin \gamma|}}\right. 1+h \left[\vphantom{\frac{A'}{|\sin \gamma|}}\right. -\text{ln}\, |\sin \gamma|  
		- \ln2 +\frac{1}{2} -\text{Ei}\left( -\frac{A'}{|\sin \gamma|} \right) \nonumber \\
		 &\qquad+\frac{4}{\pi} \langle \langle |\sin \gamma| \,\text{Ei}\left( -\frac{A'}{|\sin \gamma|} \right) \rangle \rangle_\text{i}   \left.\vphantom{\frac{A'}{|\sin \gamma|}} \right] \left.\vphantom{\frac{A'}{|\sin \gamma|}} \right\}.
	\end{align}
	% ---------------
In the regime where $A' \gtrsim 2$, both the Ei term and its double orientational average term in Eq.~(\ref{eq:exclVolNem}) essentially vanish. We see that in this regime only the parameter $h$ controls the magnitude of the twisting effect, and hence $h(\kappa D,A')$ and $D_\text{eff}(\kappa D,A')$ completely determine the system's phase behavior. However, for weakly charged rods at a low salt concentration, $A'$ can be small and the exponential integral terms in Eq.~(\ref{eq:exclVolNem}) can become important. In this case, the twisting effect not only depends on the combination $h(\kappa D,A')$, but also on $A'$ separately. Nevertheless, also in this regime it would be convenient to have a single parameter that characterizes the deviation of Eq.~(\ref{eq:exclVolNem}) from an effective hard rod-like excluded volume, $2 L^2 D_\text{eff} |\sin \gamma|$. Therefore, we define a new twisting parameter
	% ---------------
	\begin{align}\label{eq:newTwist}
		H = & 	\frac{1}{h k}\int_0^\pi   d\gamma \, \left[ \frac{E(\gamma)}{2L^2D_\text{eff} |\sin \gamma|} - 1 \right]^2\nonumber \\
		=& 	\frac{h}{k}\int_0^\pi   d\gamma \, \times \left[ \vphantom{\frac{A'}{|\sin \gamma|}}\right. -\text{ln}\, |\sin \gamma| - \ln2 +\frac{1}{2} \\
		% &\nonumber \\
		 &\left. -\text{Ei}\left( -\frac{A'}{|\sin \gamma|} \right) +\frac{4}{\pi} \langle \langle |\sin \gamma| \,\text{Ei}\left( -\frac{A'}{|\sin \gamma|} \right) \rangle \rangle_\text{i}    \right]^2 \nonumber,
	\end{align}
	% ---------------
where $k$ is a normalization factor, chosen to be
	% ---------------
	\begin{align}
		k &= \int_0^\pi\left[-\ln |\sin \gamma| - \ln2 +\frac{1}{2}\right]^2 \, d\gamma \nonumber \\
		&= \frac{\pi}{12}(3+\pi^2),
	\end{align}
	% ---------------
such that $H$ reduces to $h$ when $A' \gtrsim 2$.

 	% ---------------
	%  FIG
	% ---------------
	 	\begin{figure*}[!htb] 
			\includegraphics[width=\linewidth, bb=0 0 482 170]{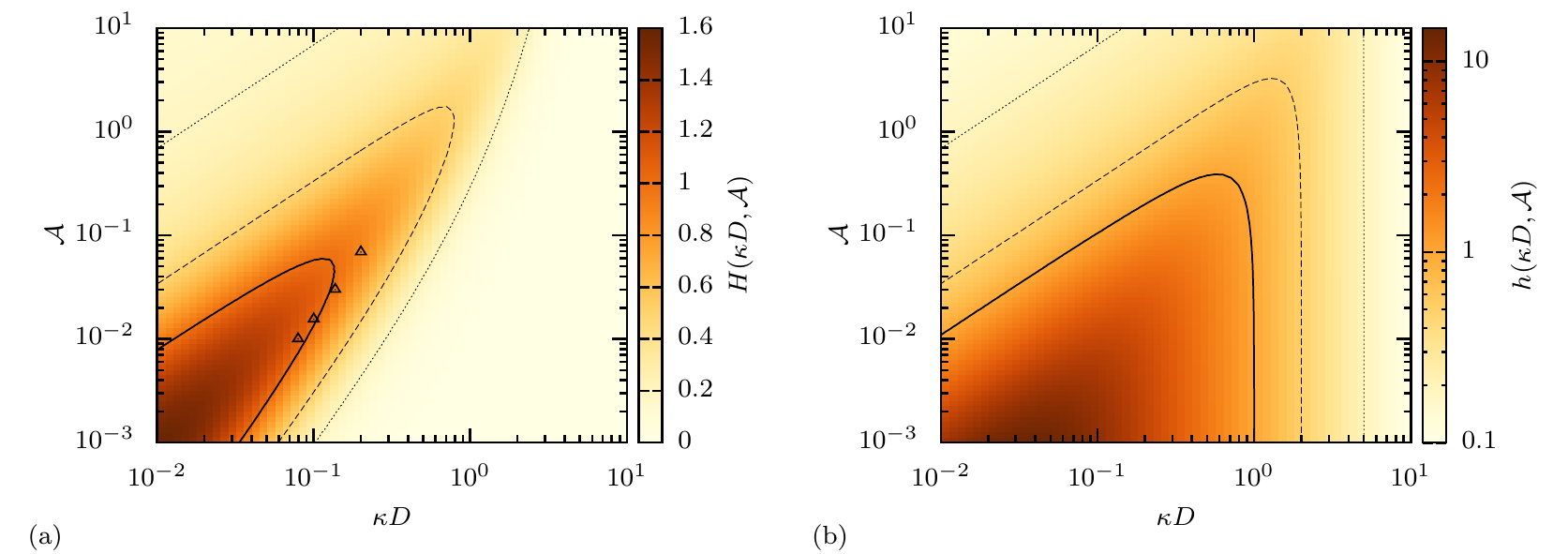}
        	\caption{Color plots indicating the value of (a) new twisting parameter $H$ (Eq.~\eqref{eq:newTwist}) with contours showing $H = 0.2,0.5,1$ (dotted, dashed, solid respectively) and (b) old twisting parameter $h$ (Eq.~\eqref{eq:oldTwist}) with contours showing $h = 0.2,0.5,1$ (dotted, dashed, solid respectively) as a function of salt concentration $\kappa D$ and charge $\mathcal{A}$, on a log-log scale. The triangles in (a) indicate the locations of the isotropic-nematic-nematic triple points found in the phase diagrams discussed following section. Note that the color bar of (b) is in a log scale.}\label{fig:contourTwist}
     \end{figure*}

In Fig.~\ref{fig:contourTwist}, we show the dependence of (a) the new twisting parameter $H$ (defined in Eq.~\eqref{eq:newTwist}) and (b) the old twisting parameter $h$ (defined in Eq.~\eqref{eq:oldTwist}) on $\kappa D$ and $\mathcal{A}$. We see that the shapes of $H$ and $h$ differ but that they agree in the upper left corner where $A'= \mathcal{A} e^{-\kappa D}/(\kappa D) \gtrsim 2$. When $A' \lesssim 2$, $h$ increases but in this parameter regime it is no longer physically relevant; interestingly, Fig.~\ref{fig:contourTwist}(a) shows that at fixed $\kappa D \lesssim 1$, the new twist parameter $H$ goes through a maximum as a function of $\mathcal{A}$ at some $\mathcal{A} \lesssim 10^{-1}$, which implies that a low (but non-zero) charge on the rods gives the strongest twisting effect.

In the following section we study the effect of twisting on the isotropic-nematic phase transition in charged rods. In Sec.~\ref{sect:ElasticConsts}, we compute the Frank elastic constants in order to see how they are influenced by the twisting effect. 

%----------------------------------------------------------------------------------------
%	Section 3
%----------------------------------------------------------------------------------------
\section{Phase diagrams}\label{sect:PhaseDiagrams}

The concentrations of the coexisting isotropic and nematic phase, $c_i$ and $c_n$ respectively, can be found using the condition that the osmotic pressures $\Pi = -\left( \partial F/\partial V \right)_{N,T}$ and chemical potentials $\mu = \left( \partial F/\partial N \right)_{V,T}$ satisfy
	% ---------------
	\begin{align}
		\Pi^\text{iso}(c_i) &= \Pi^\text{nem}(c_n) \\
		\mu^\text{iso}(c_i) &= \mu^\text{nem}(c_n).
	\end{align}
	% ---------------
 We introduce the dimensionless effective concentration
	% ---------------
	\begin{equation}
		c_\text{eff} = \frac{\pi}{4} \frac{N}{V} L^2 D_\text{eff},
		\label{eq:density}
	\end{equation}
	% ---------------
which we use rather than the usual dimensionless concentration $c=(\pi/4)(N/V)L^2D$ in order to show how twisting affects the phase behavior of charged rods. 

\begin{figure*}[htp]
			\includegraphics[width=\linewidth, bb=0 0 482 570]{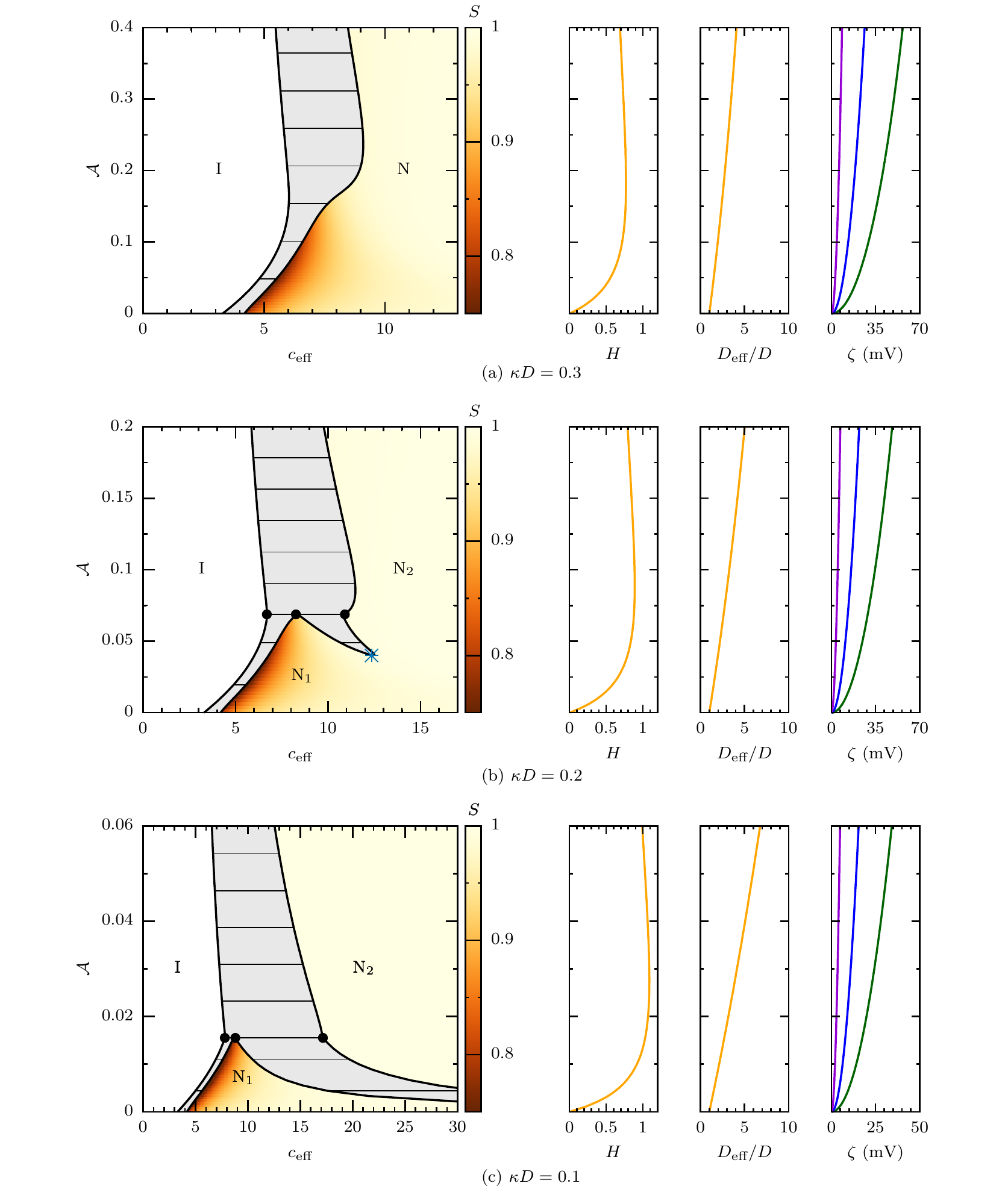}
        	\caption{Phase diagrams in the density $c_\text{eff}$ (Eq.~\eqref{eq:density})-charge $\mathcal{A}$ (Eq.~\eqref{eq:charge}) representation for (a) $\kappa D = 0.3$, (b) $\kappa D = 0.2$, and (c) $\kappa D = 0.1$, with colors showing the nematic order parameter $S$, I denoting the stable isotropic phase, N the stable nematic, N$_1$ the weakly-aligned nematic, and N$_2$ the strongly aligned nematic phase. The N$_1$-N$_2$ is critical point is denoted by an asterisk and the three coexisting phases at the triple point are denoted by black dots. The tielines that connect coexisting phases are horizontal. Next to each phase diagram, the dependence of the twisting parameter $H$ (Eq.~\eqref{eq:newTwist}) and scaled effective diameter $D_\text{eff}/D$ (Eq.~\eqref{eq:Deff}) on $\mathcal{A}$ is shown. In fourth column, the dependence of the zeta-potential $\zeta$ on $\mathcal{A}$ (in units of millivolts) is shown for diameter-Bjerrum length ratio $D/\lambda_B = 10,1,0.2$ (purple, blue, green or left to right).}
        	\label{fig:phaseDiagramsCharge}
     \end{figure*}

In Fig.~\ref{fig:phaseDiagramsCharge}, we show phase diagrams in the ($c_\text{eff}$, $\mathcal{A}$) plane for (a) $\kappa D = 0.3$, (b) $\kappa D = 0.2$, and (c) $\kappa D = 0.1$, where the horizontal tie-lines connect coexisting states and the color coding represents the nematic order parameter $S= \langle (3\cos^2\theta -1)/2 \rangle$. A first glance reveals a very rich phase diagram with isotropic-nematic and nematic-nematic coexistence, including triple points and critical points. In all three phase diagrams, we see that at $\mathcal{A}=0$ (zero charge) the expected phase transition for uncharged rods occurs, with the isotropic phase (I) existing at low concentrations, the nematic phase (N) at high concentrations, and phase coexistence between I and N in the region between $c_\text{eff}=3.29$ and $c_\text{eff}=4.19$. As we increase the charge, the twisting parameter increases and destabilizes the nematic phase, so that the I-N phase transition moves to higher effective concentrations $c_\text{eff}$. At this point, it is good to note that the definition of $c_\text{eff}$ given in Eq.~(\ref{eq:density}), involves the \emph{effective} diameter, which, as shown in Fig.~\ref{fig:contourDeff}(a), increases with increasing $\mathcal{A}$. If we were to use the concentration $c$ instead of the effective concentration, the I-N phase transition would move to \emph{lower} concentrations. We will return to this point below.

We limit the phase diagrams of Fig.~\ref{fig:phaseDiagramsCharge} to low charge, where the twisting effect is important. However, as $\mathcal{A} \to \infty$ (at fixed $\kappa D$ this corresponds to $h \to 0$), we also find a hard rod-like I-N transition, in agreement with Ref.~[\citen{Lekkerkerker1986}]. Next to each phase diagram, we show the $\mathcal{A}$-dependence of the twisting parameter $H$, the scaled effective diameter $D_\text{eff}/D$ (which is equal to the ratio $c_\text{eff}/c$), and the zeta-potential $\zeta$, i.e. the electrostatic potential on the surface of the rod as obtained from the Poisson-Boltzmann equation in a cylindrical cell (see Appendix \ref{sect:PB}). 

In Fig.~\ref{fig:phaseDiagramsCharge}(b), we see that the twisting effect is large enough to cause the nematic phase to split into a low density nematic N$_1$ and a higher density, more aligned nematic phase N$_2$. The phase diagram features a nematic-nematic (N$_1$-N$_2$) critical point and an isotropic-nematic-nematic (I-N$_1$-N$_2$) triple point. Finally, in Fig.~\ref{fig:phaseDiagramsCharge}(c), we have lowered $\kappa D$ further and we see again a triple point, and a larger region of N$_1$-N$_2$ phase coexistence, the critical point of which is outside the plotted range.

	% ---------------
	%  FIG
	% ---------------

\begin{figure*}[htp]
			\includegraphics[width=\linewidth, bb= 0 0 482 570]{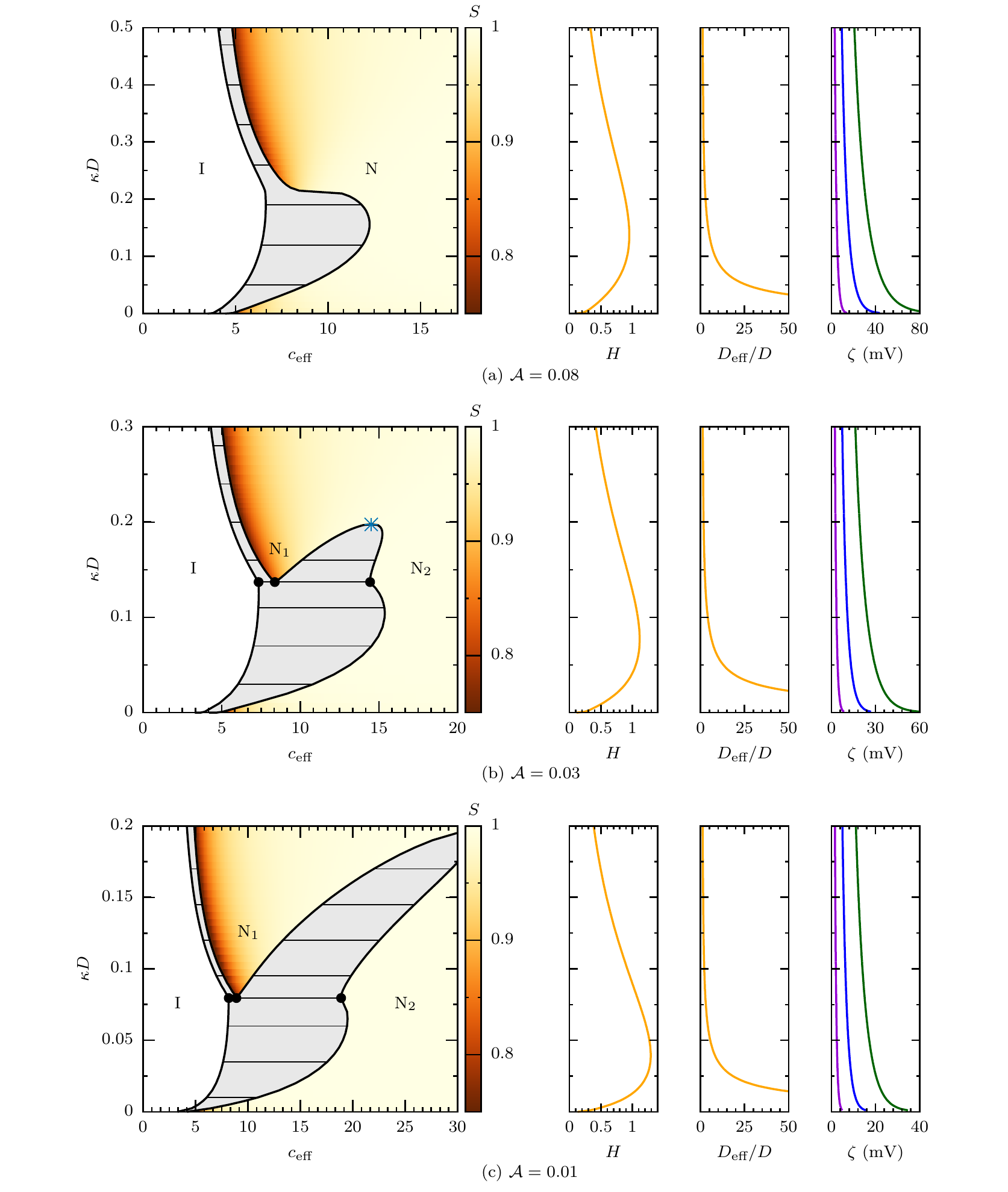}
        	\caption{Phase diagrams in the density $c_\text{eff}$-salt concentration $\kappa D$ representation for (a) $\mathcal{A} = 0.08$, (b) $\mathcal{A} = 0.03$ and (c) $\mathcal{A} = 0.01$ (see the caption of Fig.~\ref{fig:phaseDiagramsCharge} for explanation of regions and parameters), with colors showing the nematic order parameter $S$. Next to each phase diagram, the dependence of the twisting parameter $H$ and scaled effective diameter $D_\text{eff}/D$ on $\mathcal{A}$ is shown. In fourth column, the dependence of the zeta-potential $\zeta$ on $\kappa D$ (in units of millivolts) is shown for diameter-Bjerrum length ratio $D/\lambda_B = 10,1,0.2$ (purple, blue, green or left to right).}
        	\label{fig:phaseDiagramsKD}
     \end{figure*}

     In Fig.~\ref{fig:phaseDiagramsKD}, we show three phase diagrams in the ($c_\text{eff}$, $\kappa D$) plane for fixed charges characterized by (a) $\mathcal{A}=0.08$, (b)  $\mathcal{A}=0.03$, and (c)  $\mathcal{A}=0.01$. As in Fig.~\ref{fig:phaseDiagramsCharge}, we include colors showing the nematic order parameter $S$ and we plot the $\kappa D$-dependence of the twisting parameter $H$, the scaled effective diameter $D_\text{eff}/D$, and the zeta-potential $\zeta$ next to each phase diagram. Note that the effective diameter increases with decreasing $\kappa D$. We find again that the nematic phase can split into a weakly and strongly aligned nematic, when the twisting parameter $H$ is of order unity (see $H(\mathcal{A},\kappa D)$ in second columns of Figs.~\ref{fig:phaseDiagramsCharge}-\ref{fig:phaseDiagramsKD} and also triangles in Fig.~\ref{fig:contourTwist}(a) indicating locations of triple points from Figs.~\ref{fig:phaseDiagramsCharge}-\ref{fig:phaseDiagramsKD}). Given the rather arbitrary definition of $H$ (Eq.~\eqref{eq:newTwist}), one should not expect the location of the triple points to coincide exactly with the ridge of $H$ in Fig.~\ref{fig:contourTwist}(a). We stress that the phase behavior is determined by ($\mathcal{A}$, $\kappa D$) and not by the single parameter $H$.

	% ---------------
	%  FIG
	% ---------------

\begin{figure*}[htb]
			\includegraphics[width=\linewidth, bb= 0 0 482 170]{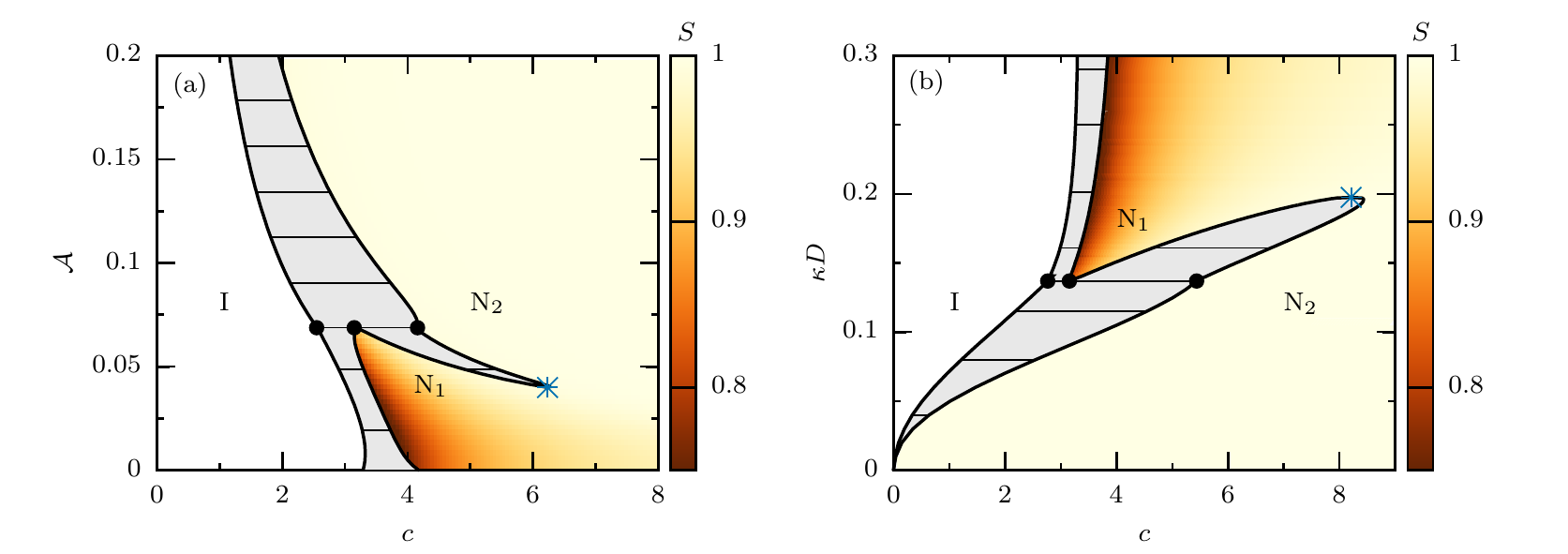}
        	\caption{Phase diagram in the (a) density $c$-charge $\mathcal{A}$ (Eq.~\eqref{eq:charge}) representation for $\kappa D = 0.2$ (compare to Fig.~\ref{fig:phaseDiagramsCharge}(b)) and (b) density $c$-salt concentration $\kappa D$ representation for $\mathcal{A} = 0.03$ (compare to Fig.~\ref{fig:phaseDiagramsKD}(b)), with colors showing the nematic order parameter $S$. The rod concentration used is $c= \frac{\pi}{4} \frac{N}{V} L^2 D$. See the caption of Fig.~\ref{fig:phaseDiagramsCharge} for explanation of regions and parameters.}\label{fig:PDConcentration}
     \end{figure*}

The phase diagrams from Figs.~\ref{fig:phaseDiagramsCharge}(b) and \ref{fig:phaseDiagramsKD}(b) are shown using the usual dimensionless rod concentration $c$ in Figs.~\ref{fig:PDConcentration}(a) and \ref{fig:PDConcentration}(b) to clarify the distinction between concentration and effective concentration discussed above. This representation makes explicit the lowering of the I-N transition densities with increasing charge and decreasing salt.

	% ---------------
	%  FIG
	% ---------------

\begin{figure*}[ht]
			\includegraphics[width=\linewidth, bb=  0 0 482 170]{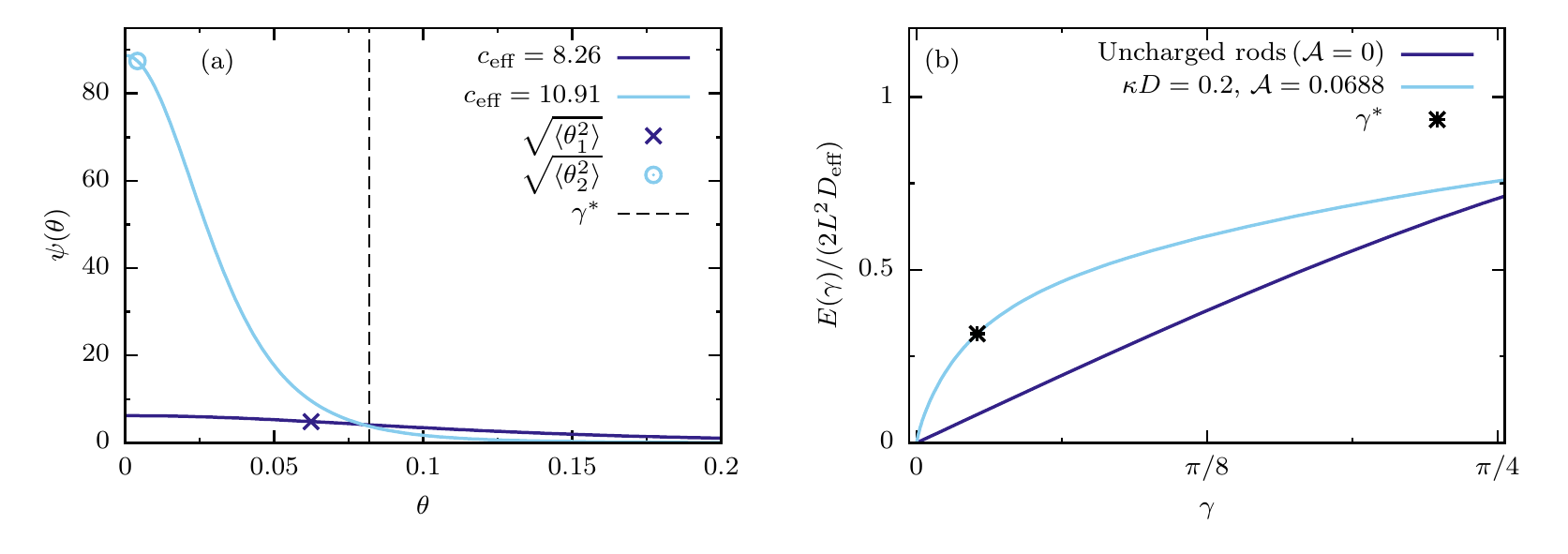}
		\caption{(a) Orientation distribution functions $\psi(\theta)$ as a function of the polar angle $\theta$ for the coexisting nematic phases at I-N$_1$-N$_2$ triple point with $\kappa D=0.2$, $\mathcal{A}=0.0688$ (see Fig.~\ref{fig:phaseDiagramsCharge}(b)). The low density nematic phase has effective concentration $c_\text{eff}=8.26$, nematic order parameter $S=0.92$, and typical angle $\sqrt{\langle \theta^2_1 \rangle} = 0.062$. The high density nematic phase has effective concentration $c_\text{eff}=10.91$, nematic order parameter $S=0.99$, and typical angle $\sqrt{\langle \theta^2_2 \rangle} = 0.0041$. The vertical dashed line denotes the cross-over angle $\gamma^*$ that seperates the ``excluded volume" $E(\gamma)$ into a steep and an essentially linear regime for $\gamma < \gamma^*$ and $\gamma > \gamma^*$, respectively. (b)~The ``excluded volume" scaled by effective volume factor $2L^2D_\text{eff}$ as a function of the angle $\gamma$ between two rods, for uncharged rods ($\mathcal{A} = 0$) and for the triple point parameters of Fig.~\ref{fig:phaseDiagramsCharge}(b), with the cross-over angle $\gamma^*$ (see text).}\label{fig:triplePt}
    \end{figure*}

In order to shed light on the microscopic origin of the charge-induced nematic-nematic demixing, we show in Fig.~\ref{fig:triplePt}(a) the orientation distribution functions for the two nematic phases at the triple point of Fig.~\ref{fig:phaseDiagramsCharge}(b) as a function of polar angle $\theta$. We can relate the existence of two nematic phases to the shape of the ``excluded volume" $E(\gamma)$ as a function of $\gamma$, the angle between two rods as shown in Fig.~\ref{fig:triplePt}(b). We can characterize the shape of $E(\gamma)$ by introducing a cross-over angle $\gamma^*$ which we give the ad hoc definition $dE(\gamma^*)/d\gamma= 4L^2D_\text{eff}$ which approximately separates $E(\gamma)$ (for small $\gamma$) into a steep part for $\gamma<\gamma^*$ and a roughly linear part for $\gamma>\gamma^*$. Two rods with polar angles $\theta$ and $\theta'$ in the less aligned nematic phase can sample a larger range of $E(\gamma)$ (note that $\gamma(\theta,\theta',\varphi-\varphi') \in [ 0,\theta+\theta']$) and often can have an angle $\gamma$ larger than $\gamma^*$. A pair of rods in the more aligned nematic phase, however, rarely has an angle $\gamma$ larger than $\gamma^*$. Therefore we can understand the appearance of the denser nematic phase as a ``condensation" in the pocket $0<\gamma<\gamma^*$; the associated loss of orientational entropy is more than compensated for by the large reduction in the excluded volume.

	% ---------------
	%  FIG
	% ---------------

%----------------------------------------------------------------------------------------
%	Section 4
%----------------------------------------------------------------------------------------
\section{Relation to experimental systems}\label{sect:experiment}

In this section, we investigate the possibility of seeing the charge-induced nematic-nematic demixing experimentally. In order for our approximations to be reliable, we require a system with $D\ll L$, $\kappa^{-1} \ll L$, and with reasonably rigid particles. We should also keep in mind that at higher densities, experimental systems of rodlike colloids undergo a nematic-smectic phase transition. For hard spherocylinders with diameter $D_\text{eff}$ and aspect-ratio $L/D_\text{eff}  > 5$ this occurs at a density approximately 47\% of the close-packed density,\cite{Bolhuis1997} which gives $c_\text{eff}=3.94$, $10.32$, $20.97$, and $42.28$ for $L/D_\text{eff} = 10$, $25$, $50$, and $100$ respectively. In other words, for sufficiently long rods the smectic phase occurs far beyond the isotropic-nematic transition. For shorter rods, or rods with a smaller effective aspect ratio, $L/D_\text{eff} \sim  4-5$, the nematic regime is small and direct isotropic-smectic transitions are to be expected.

If we look at the phase diagram with fixed $\kappa D =0.2$ (Fig.~\ref{fig:phaseDiagramsCharge}(b)) for instance, we see nematic-nematic coexistence at around $\mathcal{A} = 0.05$. In order for $D, \kappa^{-1} \ll L$  we could look at a system with $D=1$ nm, $\kappa^{-1}=5$ nm and $L \sim 100 - 1000$ nm. In water ($\lambda_B=0.7$ nm) this gives a charge density of about $v_\text{eff} = 0.11$ $e^-/$nm or surface potential $\zeta = 9.3$ mV (see Appendix \ref{sect:PB}), while in oil ($\lambda_B=8$ nm), we would need $v_\text{eff} = 0.032$ $e^-/$nm or $\zeta = 31$ mV. Similarly, for $D=5$ nm and $\kappa^{-1}=25$ nm, we would need a charge density of about $v_\text{eff} = 0.048$ $e^-/$nm or $\zeta = 4.1$ mV in water or $v_\text{eff} = 0.014$ $e^-/$nm or $\zeta = 14$ mV in oil. 

Compared to fd-virus or tobacco mosaic virus, these are very low charge densities and zeta-potentials. For example, fd virus with a length of $L=880$ nm, diameter $D=6.6$ nm, and a persistence length of 2200 nm, has about $7-10$ $e^-$/nm at room temperature with solution pH around neutral.~\cite{Tang1995, Purdy2004} For such a high charge density, the twisting effect is small ($h \lesssim 0.15$) and also not very sensitive to ionic concentration.~\cite{Tang1995} Similarly, the more rigid tobacco mosaic virus with length $L=300$ nm and diameter $D=18$ nm is very highly charged around neutral pH, with about $7-14$ $e^-$/nm.~\cite{Scheele1967,Fraden1989} 

Colloidal silica rods are another interesting model system as they are both monodisperse and rigid, but they have lower aspect ratios ($L/D \lesssim 22$) and bigger diameters ($D \gtrsim 200$ nm), making it hard to meet the conditions of small $\kappa D$ and still have $\kappa^{-1} \ll L$.~\cite{Kuijk2014, Liu2014} In Ref.~[\citen{Liu2014}], for instance, while $\kappa D \approx 0.1$, the surface potential is quite large ($\zeta \approx 70$ mV) and since the aspect ratio is low ($L/D \lesssim 5.6$) the silica rods form a plastic crystal phase rather than a nematic phase.

However, chemical modifications of fd can change its isoelectric point to be around a pH of 10, making it possible to tune the surface charge to arbitrarily small values.~\cite{Zhang2010} Ideally, a modification of fd would be found with a slightly lower isoelectric point than pH 10, such that $\kappa^{-1}$ would not be too small. Also, some polymers are rigid, have small enough diameters, and are weakly charged enough to fall in the regime of large twisting. One such a candidate is cellulose nanofibrils dispersed in water, since the surface charge density of the fibrils can be decreased to zero by lowering the pH.\cite{Fall2011} So although some degree of tuning is needed, the predicted nematic-nematic transition seems to occur in an accessible parameter regime. An issue to consider, however, is the stability with respect to irreversible aggregation due to dispersion forces; the required low charge on the rods may not be able to balance strong Van der Waals forces so some degree of index matching may be needed. 

%----------------------------------------------------------------------------------------
%	Section 5
%----------------------------------------------------------------------------------------
\section{Frank elastic constants}\label{sect:ElasticConsts}

The strong twisting effect that we identified in the low-salt and low-charge regime raises the question to what extent the uniaxial nematic phase is actually stable with respect to spontaneous twist deformations. In general the stability of bulk nematics with respect to (weak) mechanical deformations is characterized by the Frank elastic constants $K_1$ (for splaying), $K_2$ (for twisting), and $K_3$ (for bending).~\cite{deGennes, Straley1973} Mechanical stability requires all three elastic constants to be positive. In this section, we check whether or not the strong twisting effect can affect the sign of $K_i$ ($i=1,2,3$) with a focus on the twist constant $K_2$. We derive an expression for the Frank elastic constants similar to the one derived by Vroege and Odijk,~\cite{Vroege1987} which is based on the derivation for uncharged rods by Straley.~\cite{Straley1973}

In a distorted liquid crystal, the locally preferred orientation (i.e. the local nematic director) is given by $\hat{n}(\vect{r})$, where we assume that this director varies slowly in space. The relative probability of a particle at position $\vect{r}$ having orientation $\homega$ is given by the locally evaluated bulk orientation distribution function $\psi(\homega \cdot \hat{n}(\vect{r}))$. The excess free energy due to a director field distortion, up to second order in the gradients of $\hat{n}(\vect{r})$, is given in terms of the Frank elastic constants by~\cite{Straley1973}
	% ---------------
	\begin{align}\label{eq:defOfKi}
		\Delta F_d = \frac{1}{2} \int \, d \vect{r} &\left\{   K_{1} \, \left[\nabla \cdot \hat{n}(\vect{r}) \right]^2  +K_{2}\,\left[\hat{n}(\vect{r}) \cdot \nabla \times \hat{n}(\vect{r}) \right]^2 \right. \nonumber\\
		  &\quad \left. +K_{3}\,\left[\hat{n}(\vect{r}) \times \nabla \times \hat{n}(\vect{r})\right]^2 \right\}.
	\end{align}
	% ---------------
In Appendix \ref{sect:elastConstAppendix}, we show that within second-virial theory the Frank elastic constants are given by~\cite{Vroege1987,Odijk}
	% ---------------
	\begin{align}
		\label{eq:elasticConstants}
		\beta K_i D_\text{eff} = -\frac{4c_\text{eff}^2}{3\pi^2} \iint & d\homega \, d\homega' \, \left\{ \psi'(\homega \cdot \hat{n}) \psi'(\homega' \cdot \hat{n} ) \vphantom{\frac{E(\gamma)}{2 L^2 D_\text{eff}}} \right.\nonumber\\
		& \left. \times \frac{E(\gamma)}{2 L^2 D_\text{eff}}  \, F_i  \right\}, 
	\end{align}
	% ---------------
where $\psi'(\homega \cdot \hat{n})$ is a derivative of $\psi$ with respect to its argument and $F_i$ can be written in terms of local polar and azimuthal angles $\theta$ and $\phi$ as~\cite{Odijk}
	% ---------------
	\begin{align}
			\label{eq:Fi}
		&\text{Twist}: \quad &F_2 &=\frac{1}{4} \sin^3\theta \sin \theta' \cos(\phi-\phi') \nonumber \\
		&\text{Bend}: \quad &F_3 &= \cos^2\theta \sin \theta \sin \theta' \cos(\phi - \phi')\nonumber \\
		&\text{Splay}: \quad &F_1& = 3 F_2.
	\end{align}
	% ---------------

	% ---------------
	%  FIG
	% ---------------
	\begin{figure*}[htb]
			\includegraphics[width=\linewidth, bb=0 0 482 170]{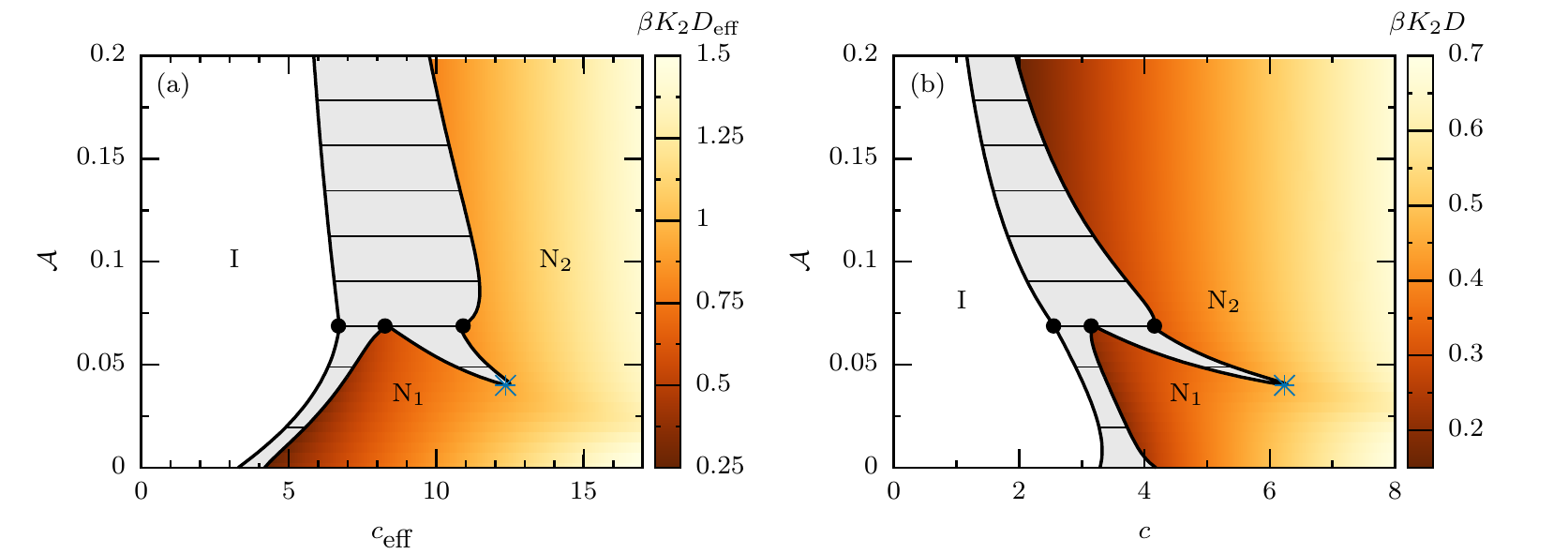}
        	\caption{Phase diagram in the density-charge $\mathcal{A}$ (Eq.~\eqref{eq:charge}) representation for $\kappa D = 0.2$ (see the caption of Fig.~\ref{fig:phaseDiagramsCharge} for explanation of regions) using (a) effective density $c_\text{eff}= \frac{\pi}{4} \frac{N}{V} L^2 D_\text{eff}$ and with colors showing dimensionless twist elastic constant $\beta K_2 D_\text{eff}$ and (b) usual dimensionless density $c= \frac{\pi}{4} \frac{N}{V} L^2 D$ and with colors showing dimensionless twist elastic constant $\beta K_2 D$.}\label{fig:K2}
     \end{figure*}

    % ---------------
	%  FIG
	% ---------------
	\begin{figure}[htb]
			\includegraphics[width=\linewidth, bb=0 0 241 170]{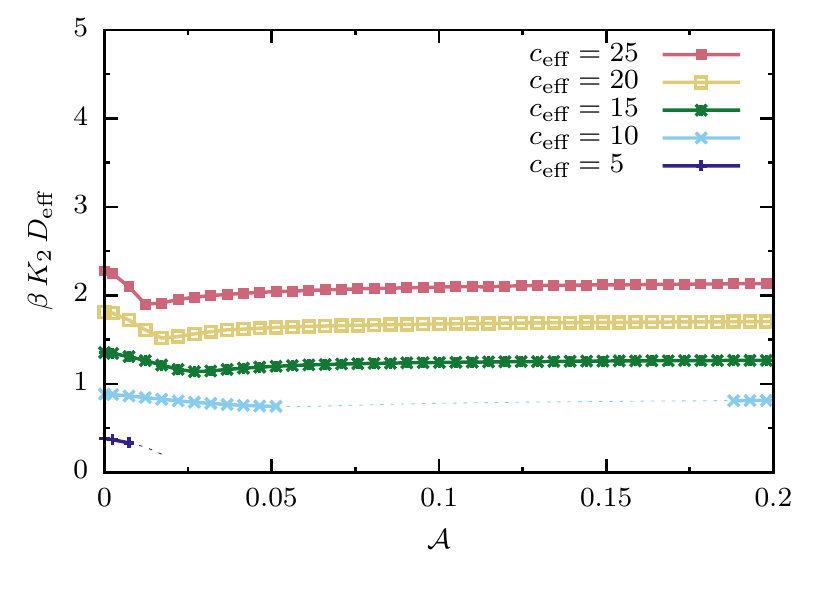}
        	\caption{Dependence of twist elastic constant $K_2$, scaled by $\beta D_\text{eff}$, on the colloidal charge parameter $\mathcal{A}$ for screening constant $\kappa D = 0.2$ and effective concentration $c_\text{eff} = 5, 10, 15, 20, 25$, from bottom to top. The dotted lines represent regions in the phase diagram (Fig.~\ref{fig:phaseDiagramsCharge}(b)) for which $c_\text{eff}$ is in the two-state coexistence gap.}\label{fig:KVsA2}
     \end{figure}

In Fig.~\ref{fig:K2}, we again show the phase diagram for $\kappa D = 0.2$ in (a) the ($c_\text{eff}$, $\mathcal{A}$) representation (see Fig.~\ref{fig:phaseDiagramsCharge}(b)) and (b) the ($c$, $\mathcal{A}$) representation (see Fig.~\ref{fig:PDConcentration}(b)), with colors now showing the twist elastic constant $K_2$ scaled in (a) by $\beta D_\text{eff}$ and (b) by $\beta D$. Note that $K_2$ is positive throughout the nematic part of the phase diagram. In Fig.~\ref{fig:KVsA2}, we show the twist elastic constant's dependence on the charge $\mathcal{A}$, for $\kappa D = 0.2$ and fixed values of effective concentration $c_\text{eff}$, which correspond to vertical lines in phase diagram Fig.~\ref{fig:K2}(a)). Here we see that the twist elastic constant has a hard-rod value for uncharged rods ($\mathcal{A}=0$), decreases for small $\mathcal{A}$ (as the twisting parameter increases), and finally increases slowly back to the hard-rod value as $\mathcal{A} \to \infty$. We see that the minimum in Fig.~\ref{fig:KVsA2} changes position slightly for different values of $c_\text{eff}$. This is because $K_2$ depends not only on the twisting effect, but also the nematic order parameter (see Fig.~\ref{fig:phaseDiagramsCharge}(b)), which first decreases and then increases with increasing $\mathcal{A}$. In addition, we calculated the bend elastic constant $K_3$ (not shown), which has a much stronger dependence on the nematic order parameter than $K_2$ does, however it is never decreased by the twisting effect. So for all parameters $\kappa D$, $\mathcal{A}$ and all nematic concentrations, we find positive Frank elastic constants.
%----------------------------------------------------------------------------------------
%	Section 6
%----------------------------------------------------------------------------------------
\section{Finite aspect-ratio charged colloidal rods}\label{sect:Simone}

In this section we investigate if spontaneous chiral symmetry breaking can occur when we consider rods of finite aspect ratio. For this purpose, we apply the recently developed second-virial density functional theory for cholesteric phases~\cite{Belli,Dussi} to a simple model of uniaxially-charged colloidal rods. This theory allows us to compute numerically the free energy $F$ as a function of the wavenumber $q$ of the chiral twist, for a given thermodynamic state of the system (e.g. at given temperature and density). We can therefore distinguish between a stable achiral nematic phase, for which the minimum of $F(q)$ is at $q=0$, a stable cholesteric phase, for which the minimum of $F(q)$ is at $q^* \neq 0$, and a spontaneous breaking of the chiral symmetry, for which the minimum of $F(q)$ is at $\pm q^* \neq 0$. As stated before, finding $K_2 \propto \frac{d^2 F(q)}{d q^2 } |_{q=0}<0$ would also be an indication that the system exhibits a spontaneous chiral symmetry breaking.

In analogy with Ref.~[\citen{Eggen}], the colloids are modeled as hard spherocylinders (HSC) of diameter $D$ and length $L$. The total charge on the rods $Z$ is fixed by embedding $N_s$ spheres interacting via a hard-core Yukawa potential (HY). The $N_s$ spheres (with $N_s$ odd) are evenly distributed along the backbone of the rod: they are separated by a distance $\delta=\frac{L}{N_s -1}$ such that two spheres are always at the extremities of the cylindrical part of the spherocylinder. The total pair potential between two charged rods is therefore
$$
U_{12} (\mathbf{r},\mathbf{\hat{\omega}},\mathbf{\hat{\omega}'}) = U_{HSC} (\mathbf{r},\mathbf{\hat{\omega}},\mathbf{\hat{\omega}'}) + \sum_{i=1}^{N_s} \sum_{j=1}^{N_s} U_{HY} (r_{ij}),
$$
where $U_{HSC}$ is the hard-core potential between spherocylinders,
$$
\beta U_{HSC} (\mathbf{r},\mathbf{\hat{\omega}},\mathbf{\hat{\omega}'})= \left\{ 
\begin{array}{ll} 
\infty & \,\, d_\text{min}(\mathbf{r},\mathbf{\hat{\omega}},\mathbf{\hat{\omega}'}) \leq D \\ 
0 & \,\, d_\text{min}(\mathbf{r},\mathbf{\hat{\omega}},\mathbf{\hat{\omega}'}) > D
\end{array} \right. ,
$$
with $d_\text{min}(\mathbf{r},\mathbf{\hat{\omega}},\mathbf{\hat{\omega}'})$ the minimum distance between two HSCs with center-of-mass separation $\mathbf{r}$ and orientations $\mathbf{\hat{\omega}},\mathbf{\hat{\omega}'}$. The sphere-sphere interaction is described by a (truncated) hard-core Yukawa potential
	%-------
	\begin{equation}
		\beta U_{HY} (r_{ij})= \left\{ 
		\begin{array}{ll} 
			\infty & \,\, r_{ij} < D \\ 
			\beta \epsilon \frac{\exp \left[ -\kappa D (r_{ij}/D - 1) \right]}{r_{ij}/D} & \,\, D \leq r_{ij}<r_\text{cut} \\ 
			0 & \,\, r_{ij} \geq r_\text{cut} 
		\end{array} \right. ,
	\end{equation}
	%-------
where $i,j$ indicates spheres belonging to rods $1,2$ respectively. The parameters $\beta \epsilon$ and $N_s$ are related by $\beta \epsilon={\left(\frac{Z}{N_s}\right)}^2$, so $N_s$ is simply a parameter that can be varied until convergence to the continuum limit is reached. As previously shown,~\cite{Eggen} this model with $N_s \geq 13$ is in excellent agreement with analytic results for the excluded volume of finite aspect-ratio rods with an effective linear charge distribution. Accordingly, we choose $N_s = 15$, which should guarantee a good agreement between the discrete-sphere and the linear-charge model. In the numerical integration we use a cutoff $r_\text{cut} \sim (1-2) $ $L$. The aspect ratio $L/D$, the total charge on the rod $Z$ and the inverse of Debye screening length $\kappa D$ are the independent physical parameters. Our approach~\cite{Belli,Dussi} relies on the numerical calculation of the excluded volume for a set of values of the chiral wavenumber $q$. Such a $q$-dependent excluded volume is calculated by performing a Monte Carlo (MC) integration using a large number of configurations and it is then used as input to calculate the free energy as a function of the chiral wavenumber $F(q)$. 

\begin{figure*}
			\includegraphics[width=\linewidth, bb=0 0 482 256]{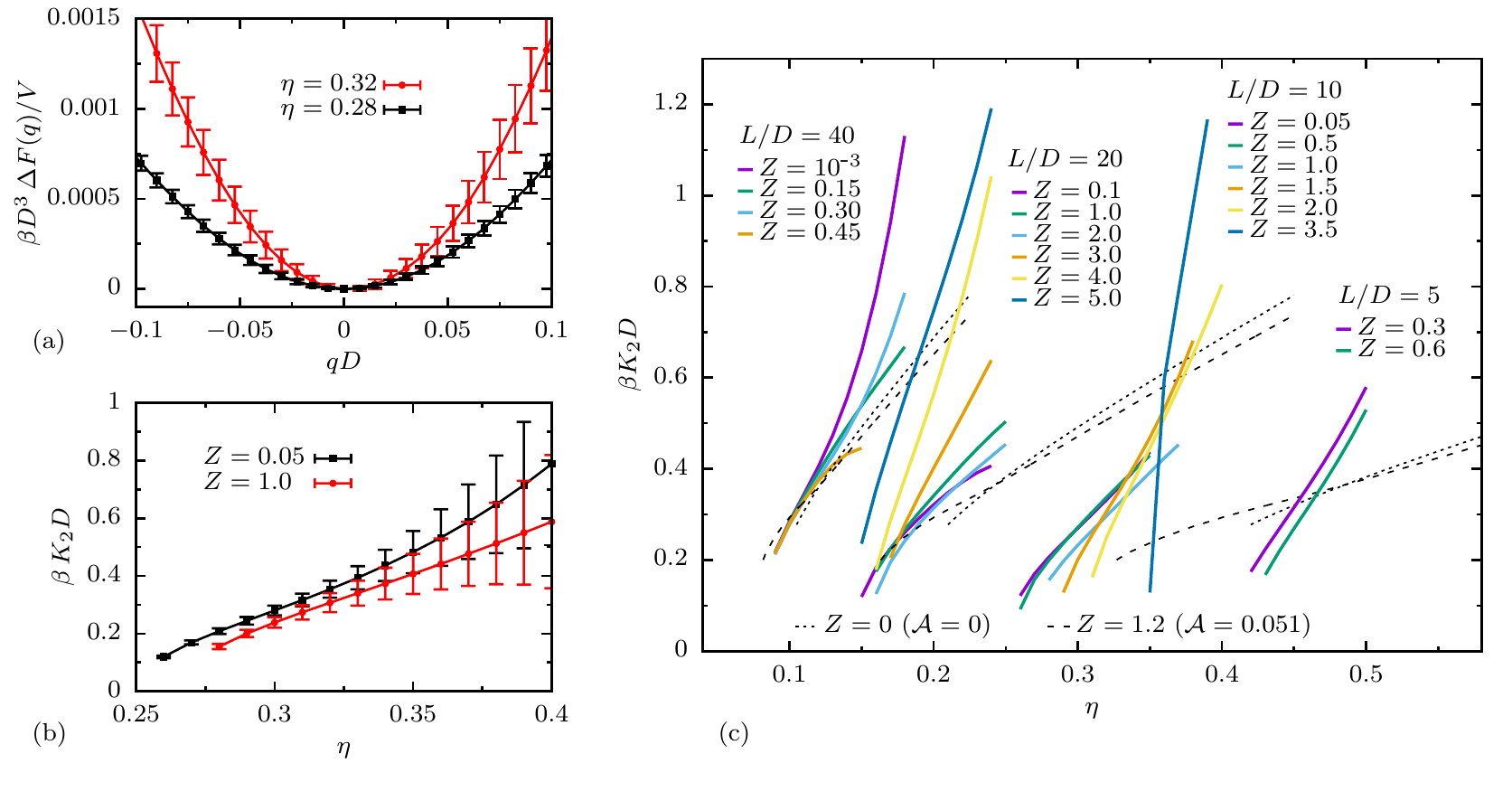}
		\caption{
		(a) Free-energy difference $\Delta F(q) =F(q)-F(q=0)$ as a function of chiral wavenumber $q D$ for two different packing fractions $\eta$, and for rods with aspect ratio $L/D=10$, total charge on rods $Z=1.0$ (corresponding to $\mathcal{A}=0.034$) divided over $N_s=15$ spheres, screening parameter $\kappa D=0.2$, and cut-off $r_\text{cut}/D=20$. The error bars are calculated by averaging over 10 independent runs of $10^{10}$ MC steps. (b) Twist elastic constant $\beta K_2 D$ calculated from second derivative of $F(q)$ as a function of packing fraction $\eta$ for $Z=0.05$ (corresponding to $\mathcal{A}=8.5 \times 10^{-5}$) and $Z=1.0$ (corresponding to $\mathcal{A}=0.034$) (for the same $N_s$, $r_\text{cut}$, and $\kappa D$ as in (a)). (c) Twist elastic constant $\beta K_2 D$ as a function of packing fraction $\eta$ for fixed screening parameter $\kappa D = 0.2$, with different aspect ratios $L/D$, and with different total charges $Z$. The solid lines are results from the MC method and the dashed lines are results from theory for $Z=0$ ($\mathcal{A}=0$) and $Z=1.2$ ($\mathcal{A}=0.051$), which are shown for aspect ratios $L/D=40$, $20$, $10$ ($L/D=5$ is outside of the plotted range).}
		\label{fig:sim}
	\end{figure*}

We investigate a few combinations of aspect ratio ($L/D$) and total charge on the rods ($Z$), with fixed screening parameter $\kappa D =0.2$, as reported in Fig.~\ref{fig:sim}. In Fig.~\ref{fig:sim}(a), we show the free-energy difference $\Delta F (q)=F(q)-F(q=0)$ as a function of chiral wavenumber $q$, for $L/D=10$, $Z = 1.0$ (corresponding to $\mathcal{A} = 0.034 $), and two different packing fractions $\eta = 0.28, \, 0.32$. In some cases, we employ different $q$-grids to check that our results are consistent. However, within our numerical accuracy no evidence of a double minimum at $q= \pm q^*\neq0$ has been observed for the entire set of parameters studied. From the second-derivative of $\Delta F (q)$ it is possible to calculate $K_2$ as a function of packing fraction $\eta$, as shown in Fig.~\ref{fig:sim}(b) for $L/D=10$ and two values of total charge $Z=0.05$ ($\mathcal{A} = 8.5 \times 10^{-5}$) and $Z=1.0$ ($\mathcal{A} = 0.034$). We see that $K_2$ increases with packing fraction and that the numerical uncertainty increases with packing fraction. In Fig.~\ref{fig:sim}(c), we show the twist elastic constant as a function of packing fraction $\eta$ for aspect ratios $L/D = 40$, $20$, $10$, $5$ and different values of total charge $Z$ on the rods. Due to the large numerical uncertainties at large packing fraction, quantitative conclusions about the actual dependence of the twist elastic constant $K_2$ on charge should be drawn carefully. However, as mentioned before, there are no indications that $K_2$ becomes negative. In addition, we show results from the previous section (i.e. Fig.~\ref{fig:K2}(b)) for total charge on the rods $Z=0$ ($\mathcal{A}=0$) and $Z=1.2$ ($\mathcal{A}= 0.051$) for aspect ratios $L/D = 40$, $20$, $10$ (the dashed curves in Fig.~\ref{fig:sim}(c)). We see that the general trend of $K_2$ is similar to that of the MC results for the largest aspect ratio $L/D=40$, but as expected, Onsager theory becomes less accurate as the aspect ratio becomes smaller. In conclusion, just as in the case of infinite rods, we do not find any evidence that a linear charge distribution can induce a spontaneous chiral symmetry breaking in colloidal rods of finite length.

    % ---------------
	%  FIG
	% ---------------

%----------------------------------------------------------------------------------------
%	Section 7
%----------------------------------------------------------------------------------------
\section{Summary and discussion}\label{sect:conclusion}

In this paper, we constructed phase diagrams for charged rods within the second-virial approximation. We found that in a low salt and low, finite charge interval, where the twisting effect dominates, a coexistence between a nematic and a second, more highly aligned nematic phase occurs as well as an isotropic-nematic-nematic triple point and a nematic-nematic critical point. The required salt and shape parameters $\kappa^{-1} \sim 5D$ and $L \gg \kappa^{-1}$ are rather easy to realize experimentally, but the required low but finite zeta-potential requires some degree of tuning near the isoelectric point.

In Refs.~[\citen{Semenov1988, Nyrkova1997}], a scaling analysis was used to treat the integral over the Mayer function in Eq.~\eqref{eq:ExclVolMayer}. Here it was predicted in a certain regime of relatively low charge density and moderate screening, that the ``excluded volume"  $E(\gamma)$ is determined by steric interactions at larger angles $\gamma$ whereas its angular dependence at small angles $\gamma$ comes from electrostatic interactions. This competition was predicted to lead to the existence of two nematic phases, one with a weak ordering and one with a very strong ordering. This is qualitatively in agreement with our findings based on full numerical evaluations. In Ref.~[\citen{Chen1996}], the nematic-nematic coexistence was confirmed to be possible in the part of the regime from Refs.~[\citen{Semenov1988, Nyrkova1997}] given by $D/L\ll \mathcal{A}/(2\pi) \ll (\kappa D)^2$ (the other part being ruled out due to many-body effects). This upper bound is indeed confirmed by our calculations. However, since we found that $\mathcal{A}/(2 \pi)$ has values at nematic-nematic coexistence in a range $\sim 0.0025-0.03$, the condition for the aspect ratio $D/L$ to be (much) smaller than this is a stricter requirement on the aspect ratio than we made in this paper, where we set it to zero from the outset. As we used the full numerical form for $E(\gamma)$ (Eq.~\eqref{eq:ExclVol}) as well as numerically solved the integral equation Eq.~\eqref{eq:EL} rather than using approximate Gaussian orientation distributions, we believe our results provide a quantitative underpinning for the nematic-nematic transitions predicted earlier in Refs.~[\citen{Semenov1988, Nyrkova1997,Chen1996}]. 

We calculated the twist elastic constant of the nematic phase of uniaxial charged rods. We showed that at a fixed effective concentration the twisting effect can reduce the twist elastic constant $K_2$, though it always remains positive. In addition, we calculated $K_2$ for uniaxial finite aspect-ratio rods, where we found no signs of negative $K_2$ either. Therefore, a uniaxial charge distribution alone seems to be not enough to break chiral symmetry, at least not within a second-virial type theory. It is an interesting possibility that by also considering the third-virial term (which includes three-body correlations), the twisting effect could be shown to stabilize a cholesteric phase. In addition, it would be interesting to see if nonlinear uniaxial charge distributions or flexibility could lead to a negative twist elastic constant. These questions are left for future studies.

%----------------------------------------------------------------------------------------
%	Section 8
%----------------------------------------------------------------------------------------
\section*{Acknowledgments}

This work is part of the D-ITP consortium, a program of the Netherlands Organization for Scientific Research (NWO) that is funded by the Dutch Ministry of Education, Culture and Science (OCW). We also acknowledge financial support from an NWO-VICI grant and from an NWO-ECHO grant.
%----------------------------------------------------------------------------------------
%	Section 9
%----------------------------------------------------------------------------------------
\appendix
\section{Poisson-Boltzmann equation}\label{sect:PB}

In order to estimate the relation between the colloidal charge $Ze$ and the zeta-potential $\zeta$, we consider a single charged rod on the symmetry axis of a cylindrical cell.~\cite{Fall2011} In the long-needle limit, we can ignore end effects, and hence the potential $\psi(r,z,\varphi)$ is not only independent of the azimuthal angle $\varphi$ but also of the Cartesian coordinate $z$ that denotes the height above the center of mass of the needle, leaving only a dependence on the radial in-plane coordinate $r$. The Poisson-Boltzmann equation for the dimensionless potential $\phi(r) = e \beta \psi(r)$ thus takes the form
\begin{align}
		\phi''(r) + \frac{1}{r} \phi'(r) &= \kappa^2 \sinh \phi(r), \quad r\geq D/2\label{eq:PB1}\\
		\phi(r \to \infty) &= 0,\label{eq:PB2}\\
		\phi'(D/2)  &= -4\pi \lambda_B \label{eq:PB3}\,\sigma,
\end{align}
where a prime denotes a derivative with respect to $r$ and where the surface charge density $e\sigma$ is given by $\sigma=Z/(\pi LD)=v_\text{eff}/(\pi D)$. Eqs.~(\ref{eq:PB1}-\ref{eq:PB3}) form a closed set that can be easily solved numerically on a discrete radial grid, and the required zeta-potential follows as $\zeta = k_\text{B} T\phi(D/2)/e$.

By taking the diameter as a unit of length, one easily checks that the Poisson-Boltzmann problem (\ref{eq:PB1})-(\ref{eq:PB3}) depends only on the dimensionless combinations $\kappa D$ and $\lambda_B v_\text{eff}$. However, in order to calculate $\zeta$ as a function of the charge parameter $\mathcal{A}= 2\pi (\lambda_B \,v_\text{eff})^2 D/\lambda_B$, one must fix $D/\lambda_B$.

%----------------------------------------------------------------------------------------
%	Section 10
%----------------------------------------------------------------------------------------
\section{Calculation of the Frank elastic constants}\label{sect:elastConstAppendix}

Within the second-virial approximation, the change in free energy due to a small spatial variation of the nematic director $\hat{n}(\vect{r})$ can be computed in terms of the Mayer function $\Phi$ as~\cite{Straley1973}
	% ---------------
	\begin{align}\label{Eq:FElastic}
		\Delta F_d 	= -\frac{1}{2}k_\text{B} T\rho^2 & \iiiint  d\vect{r} \,  d\vect{r}' \, d\homega \, d\homega'  \\
		& \times \Phi(\vect{r}'-\vect{r}; \homega, \homega')  \psi(\homega \cdot \hat{n}(\vect{r}))  \nonumber \\
		&  \times  \big[ \psi(\homega' \cdot \hat{n}(\vect{r}'))-\psi(\homega' \cdot \hat{n}(\vect{r})) \big] \nonumber.
	\end{align}
	% ---------------
 Expanding the term in square brackets to second order in derivatives of $\hat{n}$ gives
	% ---------------
	\begin{align}
		\psi(\homega'& \cdot \hat{n}(\vect{r}'))-\psi(\homega' \cdot \hat{n}(\vect{r})) =  \\
		 &\psi'(\homega' \cdot \hat{n}(\vect{r})) \left\{  \left(\vect{\xi} \cdot \nabla_r \right) \big[\hat{n}(\vect{r}) \cdot \homega'\big] \vphantom{\frac{1}{2}}\right.  \nonumber \\
		 &+\left. \frac{1}{2}(\vect{\xi} \cdot \nabla_r )^2 \big[\hat{n}(\vect{r}) \cdot \homega'\big] \right\} \nonumber\\
		& + \frac{1}{2} \psi''(\homega' \cdot \hat{n}(\vect{r})) \left\{ (\vect{\xi} \cdot \nabla_r) \big[\hat{n}(\vect{r}) \cdot \homega' \big]  \right\}^2 + \ldots,\nonumber
	\end{align}
	% ---------------
with $\vect{\xi} = \vect{r}'-\vect{r}$ and where $\psi'(\homega' \cdot \hat{n}(\vect{r}))$ is a derivative of $\psi$ with respect to its argument. Using this and integrating by parts to combine second order terms, we obtain for the free energy
	% ---------------
	\begin{align}
		\label{eq:freeEnergyElastic}
		\Delta F_d &=  -\frac{1}{2}k_\text{B} T\rho^2  \iiiint  d\vect{r} \,  d\vect{\xi} \, d\homega' \, d\homega  \, \Phi(\vect{\xi}; \homega, \homega') \nonumber\\
		& \times \Big( \psi(\homega \cdot \hat{n}(\vect{r}))   \psi'(\homega' \cdot \hat{n}(\vect{r})) \left\{ (\vect{\xi} \cdot \nabla_r)\big[  \hat{n}(\vect{r}) \cdot \homega' \big] \right\} \nonumber \\
		& -\psi'(\homega \cdot \hat{n}(\vect{r}))   \psi'(\homega' \cdot \hat{n}(\vect{r}))  \left\{(\vect{\xi} \cdot \nabla_r)\big[ \hat{n}(\vect{r}) \cdot \homega \big]\right\} \nonumber\\
		&  \left\{(\vect{\xi} \cdot \nabla_r)\big[ \hat{n}(\vect{r}) \cdot \homega' \big] \right\} \Big)  .
	\end{align}
	% ---------------
The first integral in Eq.~(\ref{eq:freeEnergyElastic}) vanishes for even Mayer functions. Now we choose to write $\vect{\xi}$ as
	% ---------------
	\begin{equation}
		\vect{\xi} =  x \frac{\homega \times \homega'} {|\homega \times \homega'|} + y \, \homega + z \, \homega' ,
	\end{equation}
	% ---------------	
where $x$ is the shortest distance between the rods as before, see Fig.~\ref{fig:spherocylinders}.

We can calculate the integral over $\vect{\xi}$ in Eq.~(\ref{eq:freeEnergyElastic}), splitting the Mayer function in two parts as

	% ---------------
	\begin{align}
		\label{eq:Mayer}
	   \Phi (x,\gamma)&=  
		\left\{
	     \begin{array}{ll}
	       \Phi_h, & x \leq D \\
	       \Phi_e & x > D 
	     \end{array}
	   \right. 
	   \\
	     &=
	   \left\{
	     \begin{array}{ll}
	       -1, & x \leq D \\
	       \text{exp}(- \beta U_\text{e}(x, \gamma)) -1, & x > D . 
	     \end{array}
	   \right. \nonumber
	\end{align}
	% ---------------
The $\vect{\xi}$ integral over hard part of Mayer function, $\Phi_h$ is then of the form~\cite{Straley1973}
	% ---------------
	\begin{align}
		\label{eq:intXiH}
		&- \int \Phi_h(\vect{\xi}; \homega, \homega') (\vect{\xi} \cdot \vect{u}) (\vect{\xi} \cdot \vect{v}) \, d \vect{\xi} \\
		&= \int_{-D}^{D}   dx  \int_{-L/2}^{L/2}  dy \int_{-L/2}^{L/2}  dz \, |\homega \times \homega' | \nonumber \\
		 &\qquad \times \left\{  y^2(\homega \cdot \vect{u})(\homega \cdot \vect{v})  + z^2(\homega' \cdot \vect{u})(\homega' \cdot \vect{v})   \vphantom{\frac{\left[ (\homega \times \homega') \cdot \vect{u}\right]\left[(\homega \times \homega') \cdot \vect{v} \right] } {| \homega \times \homega' |^{2}}} \right. \nonumber\\
		  &\qquad  \left. + x^2 \frac{\left[ (\homega \times \homega') \cdot \vect{u}\right]\left[(\homega \times \homega') \cdot \vect{v} \right] } {| \homega \times \homega' |^{2}} \right\} \nonumber \\
		&= \frac{1}{6}  L^4 D \, | \homega \times \homega' | \big[ (\homega \cdot \vect{u})(\homega \cdot \vect{v}) + (\homega' \cdot \vect{u})(\homega' \cdot \vect{v}) \big]\nonumber\\
		&\qquad + \mathcal{O}(L^2 D^3),\nonumber
	\end{align}
	% ---------------
where we introduced the shorthand notation $\vect{u} = \nabla_r\left[ \hat{n}(\vect{r}) \cdot \homega \right]$ and $\vect{v} = \nabla_r\left[ \hat{n}(\vect{r}) \cdot \homega' \right]$ and we used the fact the rods will always overlap when $-\frac{1}{2}L < y,z < \frac{1}{2}L$ and $-D<x<D$ (other overlaps are possible, but are of order $D/L$). For the electrostatic part of the Mayer function, $\Phi_e$, we calculate
	% ---------------
	\begin{align}
		\label{eq:intXiE}
		&- \int \Phi_e(\vect{\xi}; \homega, \homega') (\vect{\xi} \cdot \vect{u}) (\vect{\xi} \cdot \vect{v}) \, d \vect{\xi}  \\
		&= -2\int_{D}^{\infty}  dx  \int_{-L/2}^{L/2}  dy \int_{-L/2}^{L/2}  dz \, | \homega \times \homega' | \, \Phi_e(x, \gamma)\nonumber \\
		&\qquad \times \left\{ y^2(\homega \cdot \vect{u})(\homega \cdot \vect{v}) + z^2(\homega' \cdot \vect{u})(\homega' \cdot \vect{v})  \vphantom{\frac{\left[(\homega \times \homega') \cdot \vect{u}\right] \left[(\homega \times \homega') \cdot \vect{v}\right]} {| \homega \times \homega' |^{2}}} \right. \nonumber \\
		& \qquad + \left. x^2 \frac{\left[(\homega \times \homega') \cdot \vect{u}\right] \left[(\homega \times \homega') \cdot \vect{v}\right]} {| \homega \times \homega' |^{2}} \right\}  \nonumber \\
		&= -\frac{1}{6}  L^4  \left| \homega \times \homega' \right| \big[ (\homega \cdot \vect{u})(\homega \cdot \vect{v})+ (\homega' \cdot \vect{u})(\homega' \cdot \vect{v}) \big]\nonumber\\
		 & \qquad \qquad \times \int_{D}^{\infty}  dx \, \Phi_e(x, \gamma)  \nonumber\\
		% &\qquad+ (\homega' \cdot \vect{u})(\homega' \cdot \vect{v}) \left.\right]  \int_{D}^{\infty}  dx \, \Phi_e(x, \gamma) \nonumber\\
		 &\qquad- 2L^2 \int_{D}^{\infty}  dx \, x^2 \frac{\left[ (\homega \times \homega') \cdot \vect{u} \right] \left[(\homega \times \homega') \cdot \vect{v} \right]} {| \homega \times \homega' |} \Phi_e(x, \gamma)\nonumber\\
		&= \frac{1}{12}  L^2  E_e(\gamma)\big[ (\homega \cdot \vect{u})(\homega \cdot \vect{v}) + (\homega' \cdot \vect{u})(\homega' \cdot \vect{v}) \big]+\mathcal{O}(L^2 D^3),\nonumber
	\end{align}
where we used  
	% ---------------
	\begin{align}
		\label{eq:ExclVolE}
		E_e(\gamma) &= -2L^2 \, |\sin  \gamma| \int_D^\infty \Phi (x,\gamma)\, dx  \\
		 &= 2L^2 \kappa^{-1} \, |\sin \gamma| \left[   \ln\left( \frac{A'}{|\sin \gamma|} \right) + \gamma_E \right.\nonumber\\
		 &\quad \quad \quad \quad-\text{Ei}\left( -\frac{A'}{|\sin \gamma|}  \right)\left. \vphantom{\frac{A'}{|\sin \gamma|}} \right], \nonumber
	\end{align}
	% ---------------
and $|\homega \times \homega'| = |\sin \gamma|$. Combining Eq.~(\ref{eq:intXiH}) and (\ref{eq:intXiE}), we have
	% ---------------
	\begin{align}
			\label{eq:intXi}
		&- \int \Phi(\vect{\xi}; \homega, \homega') (\vect{\xi} \cdot \vect{u}) (\vect{\xi} \cdot \vect{v}) \, d \vect{\xi}  \\
		&= \frac{1}{12}  L^2  E(\gamma) \big[ (\homega \cdot \vect{u})(\homega \cdot \vect{v}) + (\homega' \cdot \vect{u})(\homega' \cdot \vect{v}) \big]  + \mathcal{O}(L^2 D^3).\nonumber
	\end{align}
	% ---------------	
This agrees with Ref.~[\citen{Vroege1987}]'s Eq.~(4.2) in the regime where $A' \gtrsim 2$ (in their notation $D' = E(\gamma)/(2L^2 |\sin \gamma|)$). Using Eq.~(\ref{eq:intXi}) together with Eq.~(\ref{eq:freeEnergyElastic}) and Eq.~(\ref{eq:defOfKi}), we obtain Eq.~(\ref{eq:elasticConstants}) for the elastic constants.

\bibliography{manuscript}

\end{document}